\documentclass[12pt]{article}
\usepackage{graphpap, graphics, float, tabularx, array, setspace, natbib, multirow, eurosym}
\usepackage{amsmath}
\usepackage{amstext}
\usepackage{amsbsy}
\usepackage{amsthm}
\usepackage{amsfonts}
\usepackage{amssymb}
\usepackage{amscd}
\usepackage{eucal}

\theoremstyle{plain}
\newtheorem{thm}{Theorem}[section]
\newtheorem{lem}[thm]{Lemma}

\newtheorem{ass}[thm]{Assumption}
\theoremstyle{definition}

\def\ito{It{\^o}}

\def\footnoterule{\hrule \kern2.6pt}

\def\intt{\int_0^t}
\def\intT{\int_0^T}

\def\rplu{[\,0,\infty)}

\def\as{\quad\text{\rm a.s.}}
\def\a{\alpha}
\def\b{\beta}
\def\p{\pi}
\def\l{\lambda}
\def\L{\Lambda}
\def\r{\rho}
\def\s{\sigma}
\def\k{\kappa}
\def\d{\delta}
\def\t{\tau}

\def\o{\omega}
\def\O{\Omega}

\def\m{\mu}

\def\sgn{{\rm sgn}}
\def\empty{\varnothing}
\def\0T{[\,0,T]}

\def\limT#1{\lim_{T\to\infty}\frac{#1}{T}}
\def\limt#1{\lim_{t\to\infty}\frac{#1}{t}}

\def\F{\CMcal{F}}

\long\def\symbolfootnote[#1]#2{\begingroup
\def\thefootnote{\fnsymbol{footnote}}\footnote[#1]{#2}\endgroup}
\numberwithin{equation}{section}
\newenvironment{proofLemma1}[1][Proof of Lemma \ref{totalWealthLemma}.]{\begin{trivlist}
\item[\hskip \labelsep {\bfseries #1}]}{\hfill\qed\end{trivlist}}
\newenvironment{proofLemma2}[1][Proof of Lemma \ref{localTimeLemma}.]{\begin{trivlist}
\item[\hskip \labelsep {\bfseries #1}]}{\hfill\qed\end{trivlist}}
\newenvironment{proofTheorem1}[1][Proof of Theorem \ref{wealthDistThm}.]{\begin{trivlist}
\item[\hskip \labelsep {\bfseries #1}]}{\hfill\qed\end{trivlist}}
\newenvironment{proofTheorem2}[1][Proof of Theorem \ref{wealthDistThm2}.]{\begin{trivlist}
\item[\hskip \labelsep {\bfseries #1}]}{\hfill\qed\end{trivlist}}
\theoremstyle{plain}

\renewcommand{\baselinestretch}{1.3}

\newcommand\blfootnote[1]{%
  \begingroup
  \renewcommand\thefootnote{}\footnote{#1}%
  \addtocounter{footnote}{-1}%
  \endgroup
}

\begin{document}

\centerline{\bf \LARGE{A Statistical Model of Inequality}}

\blfootnote{I would like to thank seminar participants at Princeton University, UT Austin, and the University of Houston for their helpful comments. All remaining errors are my own.}




\vskip 5pt


\vskip 25pt

\centerline{\large{Ricardo T. Fernholz\symbolfootnote[1]{Robert Day School of Economics and Finance, Claremont McKenna College, 500 E. Ninth St., Claremont, CA 91711, rfernholz@cmc.edu.}}}

\vskip 3pt

\centerline{Claremont McKenna College}

\vskip 35pt



\centerline{\large{\today}}

\vskip 70pt

\renewcommand{\baselinestretch}{1.1}
\begin{abstract}
This paper develops a nonparametric statistical model of wealth distribution that imposes little structure on the fluctuations of household wealth. In this setting, we use new techniques to obtain a closed-form household-by-household characterization of the stable distribution of wealth and show that this distribution is shaped entirely by two factors---the reversion rates (a measure of cross-sectional mean reversion) and idiosyncratic volatilities of wealth across different ranked households. By estimating these factors, our model can exactly match the U.S. wealth distribution. This provides information about the current trajectory of inequality as well as estimates of the distributional effects of progressive capital taxes. We find evidence that the U.S. wealth distribution might be on a temporarily unstable trajectory, thus suggesting that further increases in top wealth shares are likely in the near future. For capital taxes, we find that a small tax levied on just 1\% of households substantially reshapes the distribution of wealth and reduces inequality.
\end{abstract}
\renewcommand{\baselinestretch}{1.3}


\vskip 30pt

JEL Codes: E21, C14, D31

Keywords: wealth distribution, inequality, capital taxes, nonparametric methods

\vfill

\pagebreak

\section{Introduction} \label{intro}

Recent trends in income and wealth inequality have drawn much attention from both academic researchers and the general public. The detailed empirical work of \citet{Atkinson/Piketty/Saez:2011}, \citet{Davies/Sandstrom/Shorrocks/Wolff:2011}, and \citet{Piketty:2014}, among others, documents these trends for many different countries around the world and has prompted a substantive debate about their underlying causes and the appropriate policy responses, if any. The changing nature of inequality in recent decades has also raised questions about whether these trends will reverse or continue in the future.

To address these questions empirically, we develop a statistical model of inequality derived from a more general empirical approach to rank-based processes. The model features explicitly heterogeneous households that are subject to both aggregate and idiosyncratic fluctuations in their wealth holdings. In contrast to much of the related empirical literature on income dynamics (see, for example, \citealp{Browning/Ejrnaes/Alvarez:2010,Altonji/Smith/Vidangos:2013}), we impose no parametric structure on the underlying processes of household wealth accumulation and do not model or estimate these processes directly. Despite the minimal structure of our approach, we use new techniques to obtain a closed-form household-by-household characterization of the stable distribution of wealth.\footnote{In a closely related theoretical paper, \citet{Fernholz:2015a} uses these same techniques to characterize equilibrium wealth dynamics in an incomplete markets model. While this theoretical approach is necessarily less general than this paper's empirical approach, it demonstrates that our nonparametric techniques are perfectly consistent with general equilibrium.}


Our characterization of the distribution of wealth yields several new results. First, we show that the stable distribution is shaped entirely by two factors---the reversion rates and idiosyncratic volatilities of wealth for different ranked households. The reversion rates of household wealth measure the rate at which household wealth cross-sectionally reverts to the mean. Because our approach allows for wealth growth rates and volatilities that vary across different ranks in the distribution, one of this paper's contributions is to extend and generalize beyond previous work that relied on the homogeneity of Gibrat's law \citep{Gabaix:1999,Gabaix:2009}. 



Our statistical model can replicate any empirical distribution. Using the detailed new wealth shares data of \citet{Saez/Zucman:2014}, we construct such a match for the 2012 U.S. wealth distribution. According to these data, there has been a clear upward trend in top wealth shares since the mid-1980s, a fact that raises significant doubt about any model that relies on an assumed steady-state or stable distribution of wealth. One innovation of our empirical approach is that it can account for changing top wealth shares and generate estimates of the future stable distribution of wealth. We provide such empirical estimates for several different scenarios for the current underlying trend in top U.S. wealth shares. These estimates yield insight into the changing nature of inequality today. In particular, our results indicate that according to the wealth shares data of \citet{Saez/Zucman:2014}, the U.S. distribution of wealth might be on a temporarily unstable trajectory in which it splits into two divergent subpopulations, each of which forms a separate stable distribution. This unsustainable scenario suggests that further increases in the wealth shares of a tiny minority of households are likely in the near future.



The flexibility of our empirical framework allows us, in principle, to estimate the distributional effects of any tax policy. In practice, these estimates are likely to be most accurate in the case of capital taxes, since the effects of such taxes on the rate of cross-sectional mean reversion are easier to approximate. Under the assumption that a 1\% capital tax reduces the growth rate of wealth for a household paying that tax by 1\%, we show that a progressive capital tax of 1-2\% levied on just 1\% of households substantially reshapes the distribution of wealth and reduces inequality. The exact impact of this capital tax depends on the future stable U.S. distribution of wealth, but in all cases we find that this tax---which is similar to that proposed by \citet{Piketty:2014}---significantly increases the share of total wealth held by the bottom 90\% of households in the economy.

There are a number of purely empirical models of income distribution. Both \citet{Guvenen:2009} and \citet{Guvenen/Karahan/Ozkan/Song:2015}, for example, construct statistical models that replicate many aspects of the U.S. distribution of income. \citet{Browning/Ejrnaes/Alvarez:2010} and \citet{Altonji/Smith/Vidangos:2013} use indirect inference techniques to estimate detailed statistical models of household income dynamics, while \citet{Bonhomme/Robin:2010} use nonparametric techniques to analyze the different shocks that affect household earnings. In addition to the fact that this paper analyzes the distribution of wealth rather than income, an important difference between this empirical literature and our approach is that we impose minimal structure on the underlying processes of household wealth accumulation and instead model the distribution of wealth directly. We view this paper and the empirical literature on income dynamics as complements, since we focus on inequality and distributional issues and show that a parametric empirical approach is not necessary to address these issues.

There is also an extensive theoretical literature that considers the implications of both uninsurable labor income risk and uninsurable capital income risk in different macroeconomic settings. In a simple Solow growth model setting, for example, \citet{Nirei:2009} shows that introducing uninsurable investment risk yields a realistic Pareto distribution for top income and wealth shares. \citet{Jones/Kim:2014} consider a model in which entrepreneurs face heterogeneous shocks to their human capital and corresponding income, and then examine the implications of different technological and policy shocks in this setting. Adopting a more general approach, \citet{Gabaix:2009} examines several different types of stochastic processes that generate realistic steady-state Pareto distributions and that can be applied to topics ranging from the distribution of wealth to CEO compensation.\footnote{It is also possible to generate a realistic Pareto distribution of wealth in the absence of uninsurable labor and capital income risk. See, for example, \citet{Jones:2014}, who accomplishes this using a simple birth-death process combined with standard wealth accumulation dynamics.}

\citet{Benhabib/Bisin/Zhu:2011,Benhabib/Bisin/Zhu:2014} derive similarly realistic steady-state wealth distributions in a setup in which households face uninsurable investment risk and optimally choose how to consume and invest. In a closely related paper, \citet{Fernholz:2015a} uses techniques similar to this paper to replicate the U.S. wealth distribution in an environment in which rational, forward-looking households face uninsurable investment risk. The theoretical literature that considers the implications of uninsurable labor income risk is even more extensive, and includes \citet{Krussel/Smith:1998} and \citet{Castaneda/DiazGimenez/RiosRull:2003}, among others.\footnote{For a general survey of this literature, see \citet{Cagetti/DeNardi:2008}.}

This paper combines elements of these empirical and theoretical literatures on income and wealth distributions. Although we focus on the distribution of wealth, it should be noted that our techniques, results, and general approach can be applied to any rank-based system for which there is stability and some continuity. Indeed, only for unstable processes where the distribution frequently and rapidly changes is our model clearly inappropriate. In terms of the broader literature on power laws in economics and finance, then, our contribution extends previous work by \citet{Gabaix:1999,Gabaix:2009} and others who rely on equal growth rates and volatilities throughout various distributions as implied by Gibrat's law.

One of the central contributions of this paper is to construct a model that can generate an exact household-by-household match for any empirical distribution while imposing few restrictions on the underlying household wealth processes. Indeed, our approach imposes no parametric structure on either the behavior of households or the types of shocks that those households face. Furthermore, we do not assume that all households are the same ex-ante, which contrasts with many theoretical models of wealth inequality despite the empirical evidence in support of heterogeneous income profiles \citep{Guvenen:2007,Browning/Ejrnaes/Alvarez:2010}. The only assumptions that we do impose are that household wealth can be reasonably modeled by continuous semi-martingales satisfying certain basic regularity conditions and that the distribution of wealth across households is asymptotically stable. In this way, our model and results characterize the stable distribution of wealth in a more general setting than in the previous literature.

According to our characterization, the share of wealth held for each rank in the distribution depends only on the reversion rates and idiosyncratic volatilities of wealth for different ranked households. Regardless of how complex the underlying economic environment is, these two rank-based factors measure all aspects of this environment that are relevant to the stable distribution of wealth. As a consequence, the effect of any economic change on inequality can potentially be inferred from its effect on mean reversion and idiosyncratic volatility. In this way, our model provides a simple unified framework by which we may understand the distributional impact of many of the most important developments of the past few decades, such as skill-biased technical change, globalization, and changes in institutions and tax policies.

In order to match the model to the 2012 U.S. distribution of wealth, we estimate the idiosyncratic volatility of wealth for households across different ranks using previous work on the volatility of uninsurable investments \citep{Flavin/Yamashita:2002,Moskowitz/Vissing-Jorgensen:2002} and labor income \citep{Guvenen/Karahan/Ozkan/Song:2015}. With these volatility estimates, we are able to infer the implied values for the reversion rates of wealth for different ranked households. These rank-based reversion rates generate a perfect match of a stable 2012 U.S. wealth distribution.

The wealth shares data of \citet{Saez/Zucman:2014}, however, show a clear upward trend in top wealth shares starting in the mid-1980s, which is not consistent with a stable distribution. One contribution of this paper is to introduce a methodology that addresses these stability issues and can provide estimates of the future stable distribution of wealth. In other words, even though the U.S. distribution of wealth may currently be transitioning and not stable, we can still estimate where this distribution is transitioning to. One of the strengths of these estimates is that they are purely empirical and rely on no assumptions about the underlying causes of increasing inequality. In order to generate these estimates, we adjust the reversion rates for different ranked households to account for any trends in top wealth shares. Because there is substantial uncertainty about these trends, we consider several alternative scenarios for the underlying current trend in top U.S. wealth shares and estimate the trajectory of the U.S. distribution of wealth for each scenario. These estimates reveal that the future stable distribution of wealth is quite sensitive to changes in the underlying trend in top wealth shares. These estimates are, to our knowledge, the first purely empirical estimates of the changing nature of U.S. inequality.


Every alternative scenario that we consider for the underlying current trend in top U.S. wealth shares is below the rate of increasing top shares for the last few decades as reported by \citet{Saez/Zucman:2014}.\footnote{This is not true of all wealth shares data, however. Some studies based on the \emph{Survey of Consumer Finances} (SCF), for example, find smaller (or no) increases in top shares \citep{Wolff:2010}. This is one reason why we consider many different scenarios for the current trend in top shares. Furthermore, our model is easily adjusted to match any wealth shares data and any underlying trend in top shares.} The reason we do not consider a higher-trend scenario is that the rate of increasing top shares over the past few decades is difficult to reconcile with any stable distribution of wealth. In effect, our model suggests that the changes in top shares reported by \citet{Saez/Zucman:2014} might only be consistent with a divergent trajectory in which the U.S. wealth distribution separates into two subpopulations. This trajectory, in which a tiny minority of wealthy households will eventually hold all wealth, is unlikely to continue indefinitely, thus suggesting that some aspect of the economic environment is likely to change.


The result that inequality is entirely determined by two statistical factors means that, in principle, our framework can provide estimates about the effects of different tax policies on the distribution of wealth. All that is necessary to generate these estimates are the effects of these different tax policies on mean reversion and idiosyncratic volatilities of wealth for different ranked households. In practice, however, obtaining precise measurements of the differential impact of certain tax policies on households throughout the distribution of wealth is quite difficult. One important exception is the case of progressive capital taxes. The approach of this paper is uniquely suited to estimating the distributional effects of capital taxes because such taxes have more predictable effects on reversion rates. In particular, for the estimates we present in this paper, we assume that a 1\% capital tax reduces the growth rate of wealth for a household paying that tax by 1\%. Because reversion rates are measured as minus the growth rate of wealth relative to the economy for different ranked households, this 1\% capital tax will also raise the taxed household's reversion rate by 1\%. This assumption ignores any behavioral responses and distortions caused by taxes, but it is nonetheless a useful starting scenario to consider the distributional effects of progressive taxation. By adjusting the effect of capital taxes on the rank-based reversion rates, it is straightforward to extend this analysis to include any potential behavioral responses.

Using our model of the 2012 U.S. wealth distribution, we estimate the impact on inequality of a progressive capital tax of 1-2\% levied on 1\% of households in the economy. This tax is similar to the tax proposed by \citet{Piketty:2014}, and although its full effect depends on the future stable U.S. distribution of wealth, in all cases we find that this capital tax substantially reduces inequality and reshapes the distribution of wealth. Indeed, if the 2012 U.S. wealth distribution is assumed to be stable, then our estimates suggest that this tax would reduce inequality to levels comparable to those observed in the U.S. in the 1970s. We stress that this result is not a statement about total welfare and not an endorsement of a progressive capital tax. Our model does not incorporate or measure any distortions or other costs typically associated with taxes. Instead, our analysis of the distributional effects of progressive capital taxes is meant only to enhance our overall understanding of the implications of such a policy. After all, much of the recent discussion of capital taxes has focused on how they might increase government revenues or distort economic outcomes rather than how they might affect inequality and the distribution of wealth. This paper addresses this gap in our knowledge.


The rest of this paper is organized as follows. Section \ref{bench} presents the model and characterizes the stable distribution of wealth. Section \ref{empiricalApps} presents several empirical applications of the model, including an analysis of the current trajectory of the U.S. wealth distribution and an estimate of the effect of progressive capital taxes on inequality. Section \ref{conclusion} concludes. A discussion of the assumptions underlying the model and results is in Appendix \ref{assumptions}, while all of the proofs from the paper are in Appendix \ref{proofs}.



\vskip 70pt

\section{Model} \label{bench}

Consider an economy that is populated by $N > 1$ households.\footnote{For consistency and simplicity, we shall refer to households holding wealth throughout this section. However, it is important to note that our approach and results are applicable to many other empirical distributions, as mentioned in the introduction.} Time is continuous and denoted by $t \in \rplu$, and uncertainty in this economy is represented by a filtered probability space $(\O, \F, \F_t, P)$. Let $\mathbf{B}(t) = (B_1(t), \ldots, B_M(t))$, $t \in [0, \infty)$, be an $M$-dimensional Brownian motion defined on the probability space, with $M \geq N$. We assume that all stochastic processes are adapted to $\{\F_t; t \in [0, \infty)\}$, the augmented filtration generated by $\mathbf{B}$.\footnote{In order to simplify the exposition, we shall omit many of the less important regularity conditions and technical details involved with continuous-time stochastic processes.}

\subsection{Household Wealth Dynamics}
The total wealth of each household $i = 1, \ldots, N$ in this economy is given by the process $w_i$.\footnote{By considering a discrete set of explicitly heterogeneous households, this model deviates from much of the previous literature in which there is a continuum of households. This assumption is necessary for our approach and provides analytical tractability and detail in our results.} Each of these wealth processes evolves according to the stochastic differential equation
\begin{equation} \label{wealthDynamics}
 d\log w_i(t) = \m_i(t)\,dt + \sum_{z=1}^M\d_{iz}(t)\,dB_z(t),
\end{equation}
where $\m_i$ and $\d_{iz}$, $z = 1, \ldots, M$, are measurable and adapted processes. The growth rates and volatilities $\m_i$ and $\d_{iz}$ are general and practically unrestricted (they can depend on any household characteristics), having only to satisfy a few basic regularity conditions that are discussed in Appendix \ref{assumptions}. These conditions imply that the wealth processes for the households in the economy are continuous semimartingales, which represent a broad class of stochastic processes (for a detailed discussion, see \citealp{Karatzas/Shreve:1991}).\footnote{This basic setup shares much in common with the continuous-time finance literature (see, for example, \citealp{Karatzas/Shreve:1998,Duffie:2001}). Continuous semimartingales are more general than It{\^o} processes, which are common in the continuous-time finance literature \citep{Nielsen:1999}.} Indeed, the martingale representation theorem \citep{Nielsen:1999} implies that any plausible continuous wealth process can be written in the nonparametric form of equation \eqref{wealthDynamics}. Furthermore, this section's results are also likely to apply to wealth processes that are subject to sporadic, discontinuous jumps.\footnote{In a less general setting, \citet{Fernholz:2015a} presents such an extension motivated by the fact that a function with sporadic, discontinuous jumps can be approximated arbitrarily well by a continuous function.} 


The general, nonparametric structure of our approach implies that almost all previous empirical and theoretical models of income and wealth represent special cases of equation \eqref{wealthDynamics}. Indeed, most of the theoretical literature on wealth distribution assumes that households are ex-ante symmetric and hence that the growth rate parameters $\m_i$ and the standard deviation parameters $\d_{iz}$ in equation \eqref{wealthDynamics} do not persistently differ across households \citep{Benhabib/Bisin/Zhu:2011,Jones/Kim:2014,Fernholz:2015a}.\footnote{If it is assumed that households are ex-ante symmetric, however, then we can do even more with this setup. In particular, it is possible to describe the extent of economic mobility in the economy and examine the relationship between mobility and inequality. See \citet{Fernholz:2015a} for a detailed discussion.} This ex-ante symmetry is, for example, a key assumption of any analyses based on Gibrat's law \citep{Gabaix:1999,Gabaix:2009}. Even when the parameters $\m_i$ and $\d_{iz}$ do persistently differ across households, such as in much of the empirical literature on income processes \citep{Guvenen:2009,Browning/Ejrnaes/Alvarez:2010}, this heterogeneity is usually constrained by some specific parametric structure. In this sense, then, our model encompasses and extends much of the previous related literature.

One of the model's assumptions ensures that no two households' wealth dynamics are perfectly correlated over time. In other words, markets are incomplete and all households face at least some idiosyncratic risk to their wealth holdings. This assumption is consistent with both the Bewley models of uninsurable labor income risk \citep{Aiyagari:1994,Krussel/Smith:1998} and the more recent literature that considers uninsurable capital income risk \citep{Angeletos/Calvet:2006,Benhabib/Bisin/Zhu:2011,Fernholz:2015a}. This section's results characterize the effect of idiosyncratic risk to households' wealth holdings on inequality.

It is useful to describe the dynamics of total wealth for the economy, which we denote by $w(t) = w_1(t) + \cdots + w_N(t)$. In order to do so, we first characterize the covariance of wealth across different households over time. For all $i, j = 1, \ldots, N$, let the covariance process $\r_{ij}$ be given by
\begin{equation} \label{rhoIJ}
  \r_{ij}(t) = \sum_{z = 1}^M\d_{iz}(t)\d_{jz}(t).
\end{equation}
Applying \ito's Lemma to equation \eqref{wealthDynamics}, we are now able to describe the dynamics of the total wealth process $w$.

\begin{lem} \label{totalWealthLemma}
The dynamics of the process for total wealth in the economy $w$ are given by
\begin{equation} \label{totalWealthDynamics}
 d\log w(t) = \m(t)\,dt + \sum_{i=1}^N\sum_{z=1}^M\theta_i(t)\d_{iz}(t)\,dB_z(t), \as,
\end{equation}
where
\begin{equation} \label{shares}
 \theta_i(t) = \frac{w_i(t)}{w(t)},
\end{equation}
for $i = 1, \ldots, N$, and
\begin{equation} \label{mu}
 \m(t) = \sum_{i=1}^N\theta_i(t)\m_i(t) + \frac{1}{2}\left(\sum_{i=1}^N\theta_i(t)\r_{ii}(t) - \sum_{i,j=1}^N\theta_i(t)\theta_j(t)\r_{ij}(t)\right).
\end{equation}
\end{lem}

In order to characterize the stable distribution of wealth in this economy, it is necessary to consider the dynamics of household wealth by rank. One of the key insights of this model and of this paper more generally is that rank-based wealth dynamics are the essential determinants of inequality. As we demonstrate below, there is a simple, direct, and robust relationship between rank-based growth rates of wealth and the distribution of wealth. This relationship is a purely statistical result and hence can be applied to any economic environment, no matter how complex.

The first step in achieving this characterization is to introduce notation for household rank based on total wealth holdings. For $k = 1, \ldots, N$, let
\begin{equation} \label{rankWealth}
w_{(k)}(t) = \max_{1 \leq i_1 < \cdots < i_k \leq N} \min \left(w_{i_1}(t), \ldots, w_{i_k}(t)\right),
\end{equation}
so that $w_{(k)}(t)$ represents the wealth held by the household with the $k$-th most wealth among all the households in the economy at time $t$. One consequence of this definition is that
\begin{equation}
 \max (w_1(t), \ldots, w_N(t)) = w_{(1)}(t) \geq w_{(2)}(t) \geq \cdots \geq w_{(N)}(t) = \min (w_1, \ldots, w_N(t)).
\end{equation}
Similarly, let $\theta_{(k)}(t)$ be the share of total wealth held by the $k$-th wealthiest household at time $t$, so that
\begin{equation} \label{rankShares}
\theta_{(k)}(t) = \frac{w_{(k)}(t)}{w(t)},
\end{equation}
for $k = 1, \ldots, N$. 

The next step is to describe the dynamics of the household rank wealth processes $w_{(k)}$ and rank wealth share processes $\theta_{(k)}$, $k = 1, \ldots, N$. Unfortunately, this task is complicated by the fact that the max and min functions from equation \eqref{rankWealth} are not differentiable, and hence we cannot simply apply \ito's Lemma in this case. Instead, we introduce the notion of a local time to solve this problem. For any continuous process $x$, the \emph{local time} at $0$ for $x$ is the process $\L_x$ defined by
\begin{equation} \label{localTime}
 \L_x(t) = \frac{1}{2}\left(|x(t)| - |x(0)| - \intt\sgn(x(s))\,dx(s)\right).
\end{equation}
As detailed by \citet{Karatzas/Shreve:1991}, the local time for $x$ measures the amount of time the process $x$ spends near zero.\footnote{For more discussion of local times, and especially their connection to rank processes, see \citet{Fernholz:2002}.} To be able to link household rank to household index, let $p_t$ be the random permutation of $\{1, \ldots, N\}$ such that for $1 \leq i, k \leq N$,
\begin{equation} \label{pTK}
 p_t(k) = i \quad \text{if} \quad w_{(k)}(t) = w_i(t).
\end{equation}
This definition implies that $p_t(k) = i$ whenever household $i$ is the $k$-th wealthiest household in the economy, with ties broken in some consistent manner.

\begin{lem} \label{localTimeLemma}
For all $k = 1, \ldots, N$, the dynamics of the household rank wealth processes $w_{(k)}$ and rank wealth share processes $\theta_{(k)}$ are given by
\begin{equation} \label{rankWealthDynamics}
 d\log w_{(k)}(t) = d\log w_{p_t(k)}(t) + \frac{1}{2}d\L_{\log w_{(k)} - \log w_{(k + 1)}}(t) - \frac{1}{2}d\L_{\log w_{(k - 1)} - \log w_{(k)}}(t),
\end{equation}
a.s, and
\begin{equation} \label{rankWealthShareDynamics1}
  d\log\theta_{(k)}(t) = d\log\theta_{p_t(k)}(t) + \frac{1}{2}d\L_{\log\theta_{(k)} - \log\theta_{(k + 1)}}(t) - \frac{1}{2}d\L_{\log\theta_{(k - 1)} - \log\theta_{(k)}}(t),
\end{equation}
a.s., with the convention that $\L_{\log w_{(0)} - \log w_{(1)}}(t) = \L_{\log w_{(N)} - \log w_{(N+1)}}(t) = 0$.
\end{lem}

According to equation \eqref{rankWealthDynamics} from the lemma, the dynamics of wealth for the $k$-th wealthiest household in the economy are the same as those for the household that is the $k$-th wealthiest at time $t$ (household $i = p_t(k)$), plus two local time processes that capture changes in household rank (one household overtakes another in wealth) over time.\footnote{For brevity, we write $dx_{p_t(k)}(t)$ to refer to the process $\sum_{i = 1}^N1_{\{i = p_t(k)\}}dx_i(t)$ throughout this paper.} To understand this equation, note that the positive local time term $\L_{\log\theta_{(k)} - \log\theta_{(k + 1)}}$ ensures that the wealth holdings of the $k$-th wealthiest household are always larger than those of the $k+1$-th wealthiest household, and that the negative local time term $\L_{\log w_{(k - 1)} - \log w_{(k)}}$ ensures that the wealth holdings of the $k$-th wealthiest household are always smaller than those of the $k-1$-th wealthiest. Equation \eqref{rankWealthShareDynamics1} describes the similar dynamics of the rank wealth share processes $\theta_{(k)}$.

Using equations \eqref{wealthDynamics} and \eqref{totalWealthDynamics} and the definition of $\theta_i(t)$, we have that for all $i = 1, \ldots, N$,
\begin{align}
 d\log\theta_i(t) & = d\log w_i(t) - d\log w(t) \notag \\
 & = \m_i(t)\,dt + \sum_{z=1}^M\d_{iz}(t)\,dB_z(t) - \m(t)\,dt - \sum_{i=1}^N\sum_{z=1}^M\theta_i(t)\d_{iz}(t)\,dB_z(t). \label{wealthShareDynamics}
\end{align}
If we apply Lemma \ref{localTimeLemma} to equation \eqref{wealthShareDynamics}, then it follows that
\begin{equation} \label{rankWealthShareDynamics2}
\begin{aligned}
 d\log\theta_{(k)}(t) & = \left(\m_{p_t(k)}(t) - \m(t)\right)\,dt + \sum_{z=1}^M\d_{p_t(k)z}(t)\,dB_z(t) - \sum_{i=1}^N\sum_{z=1}^M\theta_i(t)\d_{iz}(t)\,dB_z(t) \\
 & \qquad + \frac{1}{2}d\L_{\log\theta_{(k)} - \log\theta_{(k + 1)}}(t) - \frac{1}{2}d\L_{\log\theta_{(k - 1)} - \log\theta_{(k)}}(t),
\end{aligned}
\end{equation}
a.s, for all $k = 1, \ldots, N$. Equation \eqref{rankWealthShareDynamics2}, in turn, implies that the process $\log\theta_{(k)} - \log\theta_{(k+1)}$ satisfies, a.s., for all $k = 1, \ldots, N - 1$,
\begin{equation} \label{rankWealthShareDynamics3}
\begin{aligned}
d\left(\log\theta_{(k)}(t) - \log\theta_{(k+1)}(t)\right) & =  \left(\m_{p_t(k)}(t) - \m_{p_t(k+1)}(t)\right)\,dt + d\L_{\log\theta_{(k)} - \log\theta_{(k + 1)}}(t) \\
& \qquad - \frac{1}{2}d\L_{\log\theta_{(k-1)} - \log\theta_{(k)}}(t) - \frac{1}{2}d\L_{\log\theta_{(k+1)} - \log\theta_{(k + 2)}}(t) \\
& \qquad + \sum_{z=1}^M\left(\d_{p_t(k)z}(t) - \d_{p_t(k+1)z}(t)\right)\,dB_z(t).
\end{aligned}
\end{equation}
The processes for relative wealth holdings of adjacent households in the distribution of wealth as given by equation \eqref{rankWealthShareDynamics3} are key to describing the stable distribution of wealth in this setup.


\subsection{Stable Distribution of Wealth}
The results presented above allow us to analytically characterize the stable distribution of wealth in this setup. Let $\a_k$ equal the time-averaged limit of the expected growth rate of wealth for the $k$-th wealthiest household relative to the expected growth rate of wealth for the whole economy, so that
\begin{equation} \label{alphaK}
 \a_k = \limT1\intT\left(\m_{p_t(k)}(t) - \m(t)\right)\,dt,
\end{equation}
for $k = 1, \ldots, N$. The relative growth rates $\a_k$ determine the reversion rates of household wealth and are a rough measure of the rate at which wealth cross-sectionally reverts to the mean. These parameters incorporate all aspects of the economic environment, including taxes and diminishing returns to wealth accumulation.

In a similar manner, we wish to define the time-averaged limit of the volatility of the process $\log\theta_{(k)} - \log\theta_{(k + 1)}$, which measures the relative wealth holdings of adjacent households in the distribution of wealth. For all $k = 1, \ldots, N - 1$, let $\s_k$ be given by
\begin{equation} \label{sigmaK}
 \s^2_k = \limT1\intT\sum_{z=1}^M\left(\d_{p_t(k)z}(t) - \d_{p_t(k+1)z}(t)\right)^2\,dt.
\end{equation}
The relative growth rates $\a_k$ together with the volatilities $\s_k$ entirely determine the shape of the stable distribution of wealth in this economy. Finally, for all $k = 1, \ldots, N$, let
\begin{equation} \label{kappa}
 \k_k = \limT1\L_{\log\theta_{(k)} - \log\theta_{(k + 1)}}(T).
\end{equation}
Let $\k_0 = 0$, as well. In Appendix \ref{proofs}, we show that the parameters $\a_k$ and $\k_k$ are related by $\a_k - \a_{k+1} = \frac{1}{2}\k_{k-1} - \k_k + \frac{1}{2}\k_{k+1}$, for all $k = 1, \ldots, N-1$. 

The distribution of wealth in this economy is \emph{stable} if the limits in equations \eqref{alphaK}-\eqref{kappa} all exist and if the limits in equations \eqref{sigmaK}-\eqref{kappa} are positive constants. Throughout this paper, we assume that the limits in equations \eqref{alphaK}-\eqref{kappa} do in fact exist.

The \emph{stable version} of the process $\log\theta_{(k)} - \log\theta_{(k + 1)}$ is the process $\log\hat{\theta}_{(k)} - \log\hat{\theta}_{(k + 1)}$ defined by
\begin{equation} \label{stableVersion}
  d\left(\log\hat{\theta}_{(k)}(t) - \log\hat{\theta}_{(k+1)}(t)\right) = -\k_k\,dt + d\L_{\log\hat{\theta}_{(k)} - \log\hat{\theta}_{(k + 1)}}(t) + \s_k\,dB(t),
\end{equation}
for all $k = 1, \ldots, N - 1$.\footnote{For each $k = 1, \ldots, N$, equation \eqref{stableVersion} implicitly defines another Brownian motion $B(t)$, $t \in [0, \infty)$. These Brownian motions can covary in any way across different $k$.} The stable version of $\log\theta_{(k)} - \log\theta_{(k+1)}$ replaces all of the processes from the right-hand side of equation \eqref{rankWealthShareDynamics3} with their time-averaged limits, with the exception of the local time process $\L_{\log\theta_{(k)} - \log\theta_{(k + 1)}}$. By considering the stable version of these relative wealth holdings processes, we are able to obtain a simple characterization of the distribution of wealth.


\begin{thm} \label{wealthDistThm}
There is a stable distribution of wealth in this economy if and only if $\a_1 + \cdots + \a_k < 0$, for $k = 1, \ldots, N - 1$. Furthermore, if there is a stable distribution of wealth, then for $k = 1, \ldots, N - 1$, this distribution satisfies
\begin{equation} \label{wealthDistEq}
 \limT1\intT\left(\log\hat{\theta}_{(k)}(t) - \log\hat{\theta}_{(k + 1)}(t)\right)\,dt = \frac{\s^2_k}{-4(\a_1 + \cdots + \a_k)}, \as
\end{equation}
\end{thm}

Theorem \ref{wealthDistThm} provides an analytic household-by-household characterization of the entire stable distribution of wealth. This is achieved despite minimal assumptions on the processes that describe the dynamics of household wealth over time. In fact, as the time-averaged limits in equations \eqref{alphaK}-\eqref{kappa} and \eqref{wealthDistEq} suggest, we do not assume that a steady-state distribution of wealth even exists, but rather that the system of wealth processes is asymptotically stable in the sense that the limits \eqref{alphaK}-\eqref{kappa} exist. Furthermore, as long as the relative growth rates, volatilities, and local times that we take limits of do not change drastically and frequently over time, then the distribution of the stable versions of $\theta_{(k)}$ from Theorem \ref{wealthDistThm} will accurately reflect the distribution of the true versions of these rank wealth share processes.\footnote{\citet{Fernholz:2002} discusses these issues in more detail and shows that Theorem \ref{wealthDistThm} provides an accurate depiction of the U.S. stock market.} For this reason, we shall assume that equation \eqref{wealthDistEq} approximately describes the true versions of $\theta_{(k)}$ throughout much of this paper. The issue of stability of the distribution of wealth is discussed in more detail in Section \ref{empiricalApps}.


The theorem yields two important insights. First, it shows that an understanding of rank-based household wealth dynamics is sufficient to describe the entire distribution of wealth. It is not necessary to directly model and estimate household wealth dynamics by name, denoted by index $i$, as is common in the literature on earnings dynamics. Second, the theorem shows that the only two factors that affect the distribution of wealth are the rank-based reversion rates, measured by the quantities $-\a_k$, and the rank-based volatilities, $\s_k$. To understand the effect of policy, institutions, technology, globalization, or any other relevant factor on inequality, then, it is necessary only to understand their effect on these reversion rates and volatilities. Furthermore, if quantitative estimates of these effects can be obtained, then Theorem \ref{wealthDistThm} provides a quantitative description of the impact on inequality. This observation underlies our analysis of the effect of progressive capital taxes on the distribution of wealth in Section \ref{empiricalApps}.

The characterization in equation \eqref{wealthDistEq} extends earlier analyses of power law distributions based on Gibrat's law. Indeed, Gibrat's law is a very special case of Theorem \ref{wealthDistThm} in which the cross-sectional mean reversion and volatility parameters $\a_k$ and $\s_k$ are equal across different ranks $k$. In this case, the theorem confirms that in fact this setup yields a Pareto distribution (a straight line in a log-log plot of rank $k$ versus wealth holdings $\theta_k$) as in \citet{Gabaix:1999,Gabaix:2009}. One of this paper's contributions is to move past Gibrat's law and characterize how growth rates and volatilities that vary across different ranks can generate any empirical distribution. In Section \ref{empiricalApps}, we use this flexibility to construct an exact match of the U.S. wealth distribution.


According to Theorem \ref{wealthDistThm}, asymptotic stability of the distribution of wealth requires that the reversion rates $-\a_k$ must sum to positive quantities, for all $k = 1, \ldots, N - 1$. Stability, then, requires a mean reversion condition in the sense that the growth rate of wealth for the wealthiest households in the economy must be strictly below the growth rate of wealth for less wealthy households. As a consequence, even if some households are more skilled than others in that, all else equal, their expected growth rates of wealth $\m_i(t)$ are higher, stability still requires that these skilled households have lower expected growth rates of wealth when they occupy the upper ranks of the wealth distribution.

If the mean reversion condition from Theorem \ref{wealthDistThm} is not satisfied, then the distribution of wealth will separate into divergent subpopulations. In order to describe this unstable scenario, for $1 \leq m \leq N$, let
\begin{equation} \label{Am}
 A_m = \frac{\a_1 + \cdots + \a_m}{m},
\end{equation}
so that $A_m$ is the average relative growth rate of wealth for the top $m$ wealthiest households in the economy.

\begin{thm} \label{wealthDistThm2}
Suppose that the reversion rates are such that $\a_1 + \cdots + \a_k \geq 0$, for some $k < N$, and that there exists some $m < N$ such that
\begin{equation}
 A_m = \max_{1 \leq k \leq N} A_k \qquad \text{and} \qquad A_m > A_l \;\; \text{for $l \neq m$.}
\end{equation}
In this case, there exists a stable distribution of wealth for the subset of ranked households $w_{(1)}, \ldots, w_{(m)}$, and the share of total wealth held by this top subset of households satisfies\footnote{It is unlikely but possible that there are two or more maxima and hence $A_m = A_l$ for some $l \neq m$. In this case, there still exists a divergent subset of households, although equation \eqref{divergeDesc} must be changed to a time-averaged limit. This divergent subset is made up of the smallest subset of households with an average relative growth rate that is a maximum. See \citet{Fernholz/Fernholz:2014} for a proof.}
\begin{equation} \label{divergeDesc}
 \lim_{T\to\infty}\theta_{(1)}(T) + \cdots + \theta_{(m)}(T) = 1, \as
\end{equation}
\end{thm}

The stable distribution of wealth for the top subpopulation of households $w_{(1)}, \ldots, w_{(m)}$ is described using Theorem \ref{wealthDistThm}, with the parameters $\a_k$ defined as the time-averaged limit of the growth rates of wealth for different ranked households relative to the growth rate of the total wealth held by this group of households (same as equation \eqref{alphaK}, but with $\m(t)$ replaced by the growth rate of $w_{(1)} + \cdots + w_{(m)}$) and the volatility parameters $\s_1, \ldots, \s_{m - 1}$ unchanged. This wealthy subset of households forms a separate stable distribution and eventually holds all wealth in the economy. During this process of divergence, this top subset of households gradually separates from the rest of the population so that eventually there is no more mobility between groups.\footnote{See the proof of Theorem \ref{wealthDistThm2} in Appendix \ref{proofs} for a proof of this mobility result.}

As we shall explain in Section \ref{empiricalApps}, the divergent scenario of Theorem \ref{wealthDistThm2} may in fact be relevant to the current trajectory of the U.S. distribution of wealth. This theorem describes a particularly blatant form of divergence in which the wealth holdings of some subset of rich households is growing more quickly than the total wealth of the economy. In fact, the distribution of wealth is unstable even if all households have equal growth rates of wealth and hence no subset of households is growing faster than any other. In terms of the rank-based relative growth rates $\a_k$, this implies that $\a_1 = \cdots = \a_N$ and hence that all reversion rates are equal to zero. This special case has been analyzed in detail by both \citet{Fargione/Lehman/Polasky:2011} and \citet{Fernholz/Fernholz:2014}, the latter of who show that in this scenario wealth becomes increasingly concentrated over time in the sense that the time-averaged limit of the top wealth share $\theta_{(1)}$ converges to one.


\vfill
\pagebreak

\section{Empirical Applications} \label{empiricalApps}

We wish to use the empirical approach of Section \ref{bench} for several applications. One of this approach's strengths is that it can replicate any empirical distribution. In this section, we estimate the nonparametric model using the detailed new U.S. wealth distribution data of \citet{Saez/Zucman:2014}.\footnote{We use the wealth shares data of \citet{Saez/Zucman:2014} because of its great detail, especially for top shares. It should be noted, however, that the procedure of estimating the model described in this section can be applied to any distribution of wealth.} We then use this estimated model to analyze future trends in inequality and to consider the distributional effects of progressive capital taxes under different assumptions about the future.

\subsection{Estimating the Model} \label{usWealthDist}
Throughout this paper, we set the number of households in the economy $N$ equal to one million. This number balances the need for realism with the need to perform computations and simulations in a reasonable amount of time. Furthermore, all of our results are essentially unchanged with an even larger number of households in the economy.

According to equation \eqref{wealthDistEq} from Theorem \ref{wealthDistThm}, the stable distribution of wealth in the economy satisfies, for all $k = 1, \ldots, N - 1$,
\begin{equation} \label{wealthDistEq2}
 \limT1\intT\left(\log\hat{\theta}_{(k)}(t) - \log\hat{\theta}_{(k + 1)}(t)\right)\,dt = \frac{\s^2_k}{-4(\a_1 + \cdots + \a_k)}, \as
\end{equation}
This equation establishes a simple relationship between inequality as measured by the time-averaged limit of $\log\hat{\theta}_{(k)}(t) - \log\hat{\theta}_{(k + 1)}(t)$, the rank-based reversion rates, $-\a_k$, and the rank-based volatilities, $\s_k$. As discussed in Section \ref{bench}, we shall assume that equation \eqref{wealthDistEq2} approximately describes the true versions of the wealth shares $\theta_{(k)}$. Ideally, we would use detailed panel data on individual households' wealth holdings over time to estimate the quantities $\a_k$ and $\s_k$, and then confirm that these estimates replicate the observed wealth shares $\theta_{(k)}$.\footnote{This is the approach of \citet{Fernholz:2002}, who shows that a similar model accurately replicates the distribution of total market capitalizations for U.S. stocks.} Of course, a comprehensive panel data set on household wealth holdings in the U.S. does not yet exist. Given these data limitations, we instead choose to use estimates of the wealth shares $\theta_{(k)}$ and the rank-based volatilities $\s_k$ to infer the values of the rank-based reversion rates $\a_k$ via equation \eqref{wealthDistEq2}. 


The first step in this process is to generate estimates of the rank-based volatilities $\s_k$. According to equation \eqref{sigmaK}, these volatilities correspond to the time-averaged limit of the quadratic variation for the process $\log\theta_{(k)} - \log\theta_{(k+1)}$, which measures the relative wealth holdings of households that are adjacent in the wealth distribution. There is no research that directly estimates this quantity for U.S. household wealth holdings, but there is research that estimates the volatility of labor income and of the idiosyncratic component of capital income. We shall use these estimates to construct estimates of the volatility of relative wealth holdings.


In order to generate these estimates, consider the dynamic relationship between household wealth holdings, capital income, labor income, and consumption. If we let $\l_i, c_i,$ and $r_i$ denote, respectively, the after-tax labor income, consumption, and after-tax return processes for household $i = 1, \ldots, N$, then the dynamics of wealth over time for each household $i$ are given by
\begin{equation}
 dw_i(t) = w_i(t)r_i(t)\,dt + \left(\l_i(t) - c_i(t)\right)\,dt = w_i(t)\left(r_i(t) + \frac{\l_i(t) - c_i(t)}{w_i(t)}\right)\,dt.
\end{equation}
It follows that to estimate the volatility of log household wealth holdings, we need estimates of the volatility of both idiosyncratic after-tax investment returns and idiosyncratic fluctuations in after-tax labor income minus consumption (savings), with the latter volatility expressed relative to total wealth holdings. Ownership of primary housing and private equity are well-documented examples of uninsurable investments subject to idiosyncratic risk. Following \citet{Angeletos:2007}, \citet{Benhabib/Bisin/Zhu:2011}, \citet{Fernholz:2015a}, and much of the growing macroeconomic literature to feature idiosyncratic capital income risk, we set the standard deviation of idiosyncratic investment returns equal to $0.2$. This value is derived from the empirical analyses of \citet{Flavin/Yamashita:2002} and \citet{Case/Shiller:1989} for ownership of primary housing, and \citet{Moskowitz/Vissing-Jorgensen:2002} for private equity.

Measuring the volatility of idiosyncratic fluctuations in after-tax labor income minus consumption relative to total wealth holdings is more difficult. Indeed, there is no research that directly measures this volatility. For our purposes, we wish to construct low and high estimates of this volatility, which, because it depends on household wealth holdings, will vary across the distribution of wealth. To construct these estimates, we first follow \citet{Guvenen/Karahan/Ozkan/Song:2015} and set the standard deviation of changes in log labor income equal to $0.5$. We then combine this figure with the earnings and wealth holdings data from the 2007 \emph{Survey of Consumer Finances} as reported by \citet{DiazGimenez/Glover/RiosRull:2011} to construct estimates of the volatility of labor income relative to wealth holdings, which we assume is equal to the volatility of idiosyncratic fluctuations in after-tax labor income minus consumption relative to total wealth. Intentionally, these estimates may overstate the true volatility since they assume that all fluctuations in labor income both are idiosyncratic and lead to corresponding fluctuations in labor income minus consumption (there is no offsetting change in consumption).


If we add the estimated standard deviation of idiosyncratic fluctuations in labor income minus consumption relative to total wealth holdings to our estimate of the standard deviation of idiosyncratic investment returns $0.2$, then we obtain high estimates of the rank-based volatilities $\s_k$.\footnote{More precisely, these estimates are generated by adding the estimated variance of idiosyncratic fluctuations in labor income minus consumption relative to total wealth holdings to the estimated variance of idiosyncratic investment returns, multiplying by two, and then taking the square root.} These high estimates are reported in the third column of Table \ref{sigmasTab}. We shall also consider low estimates of $\s_k$, in which we assume that there is no volatility of idiosyncratic fluctuations in labor income minus consumption so that the rank-based volatilities are all equal to $\sqrt{0.2^2 + 0.2^2} = 0.28$. These low estimates are reported in the second column of Table \ref{sigmasTab}.


Taken together, the low and high estimates of $\s_k$ cover a wide range of plausible values for the rank-based volatilities.\footnote{One implication of this is that these estimates of $\s_k$ also imply a range of plausible values for the reversion rates $-\a_k$, since these values are inferred using estimates of $\s_k$ and the wealth shares $\theta_{(k)}$.} This wide range reflects the substantial uncertainty that exists regarding the true volatility of the process $\theta_{(k)} - \theta_{(k+1)}$, which measures the relative wealth holdings of households that are adjacent in the wealth distribution. Despite this uncertainty, however, these low and high estimates of $\s_k$ very likely provide lower and upper bounds for the true values of these parameters. Indeed, all of the available empirical evidence suggests that the true values of $\s_k$ are above our low estimates and below our high estimates. Future work that estimates these rank-based volatilities more accurately will help to narrow this range.

The last step in estimating the model and matching the U.S. wealth distribution is to infer values for cross-sectional mean reversion $-\a_k$ using equation \eqref{wealthDistEq2}. Normally, this is straightforward since the system of $N - 1$ equations \eqref{wealthDistEq2} together with the fact that $\a_1 + \cdots + \a_N = 0$ yields a solution. The problem, in this case, is that there are no wealth shares data that report the wealth holdings of each individual household in the economy $\theta_k$. Indeed, the data of \citet{Saez/Zucman:2014} report the wealth holdings of just a few subsets of U.S. households.

\begin{figure}[p]
\begin{center}
\vspace{-30pt}
\hspace{-15pt}\scalebox{.63}{ {\includegraphics{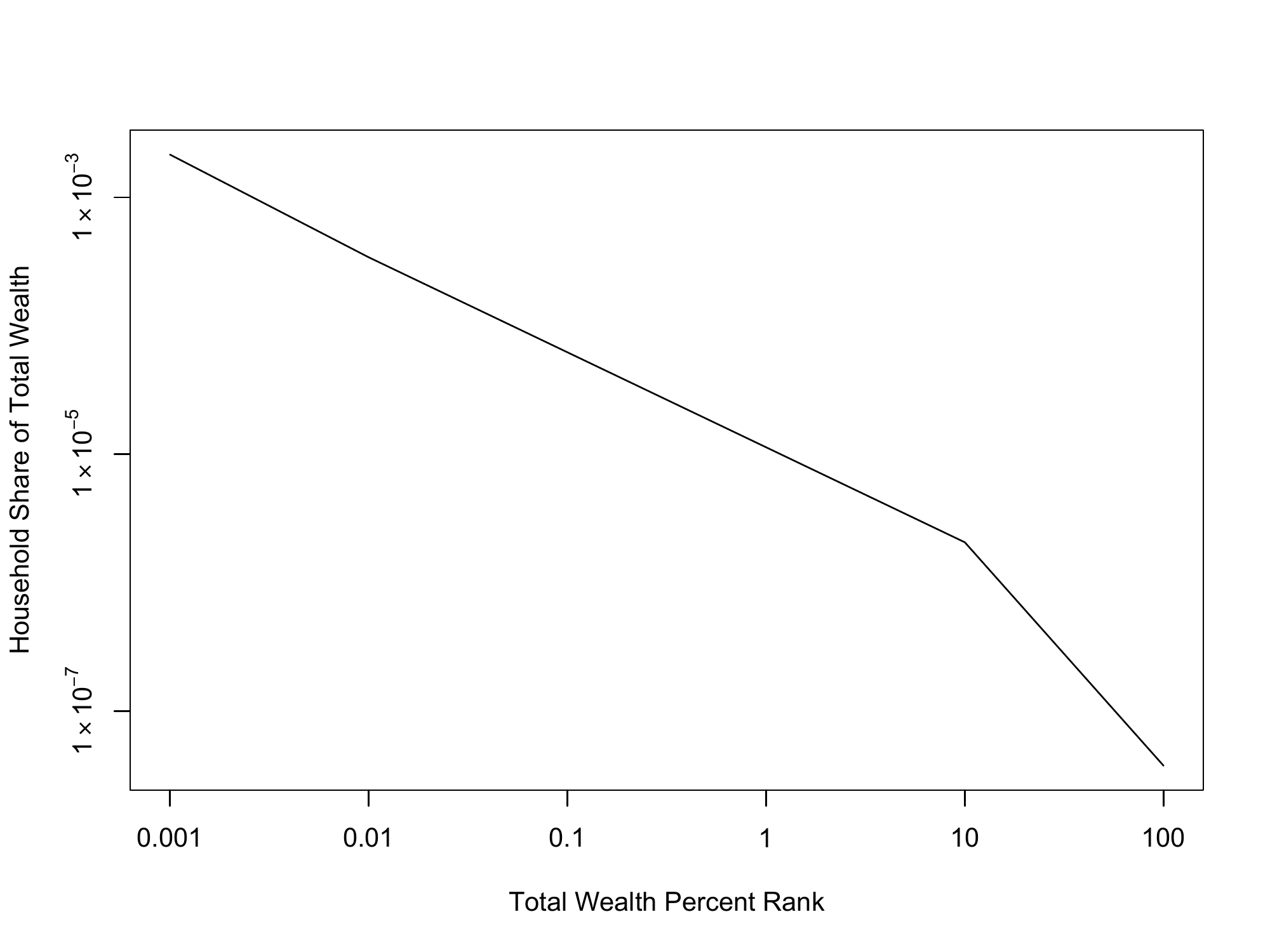}}}
\end{center}
\vspace{-24pt} \caption{Household wealth shares for the model matched to the U.S. wealth distribution in 2012.}
\label{wealthBasicFig}
\end{figure}

To fill in the missing wealth shares data, we assume a Pareto-like distribution of wealth in which the parameter of the Pareto distribution varies across different subsets of households in a way that matches the data. In fact, we find that varying the Pareto parameter across just three subsets of households achieves a nearly perfect match of the 2012 U.S. wealth distribution as reported by \citet{Saez/Zucman:2014}. Changing the Pareto parameter in this way is equivalent to assuming that the log-log plot of household rank versus household wealth holdings consists of three connected straight lines with different slopes.\footnote{More specifically, there is one straight line for the top 0.01\% of households, that line connects to another straight line with a different slope for the top 0.01-10\% of households, and that line connects to a third straight line with a different slope for the bottom 90\% of households. Such a distribution generates a total absolute error relative to the true U.S. distribution of wealth in 2012 of just over 0.5\%. While it is certainly possible to vary this slope across even more subsets of households, our approach balances simplicity and accuracy without altering the model's basic results or predictions.} A plot of this kind that achieves the closest possible match for the 2012 U.S. wealth distribution is shown in Figure \ref{wealthBasicFig}. This plot shows the value of log wealth shares $\theta_{(k)}$ versus the log of rank $k$. Once the household wealth shares $\theta_k$ are set, the rank-based reversion rates $-\a_k$ are inferred by solving the system of $N - 1$ equations \eqref{wealthDistEq2}.

In the case of a standard Pareto distribution, a log-log plot as in Figure \ref{wealthBasicFig} appears as a single straight line with slope equal to the inverse of the Pareto parameter. Our approach is slightly more general and is preferred to restricting the stable distribution of wealth to a distribution such as Pareto or lognormal since it allows the model to more closely replicate the empirical distribution of wealth. This increased accuracy and flexibility highlights one of the model's advantages. Furthermore, our basic qualitative results remain unchanged even if we do restrict the stable distribution of wealth to a common distribution.

\subsection{The U.S. Wealth Distribution, Present and Future} \label{usWealthPresentFuture}
The process of estimating the model as described in the previous subsection can be applied to any empirical distribution of wealth. This process yields implied values for the rank-based reversion rates $-\a_k$ using wealth shares data and estimates of the volatilities $\s_k$. Thus, if we use the 2012 U.S. wealth shares data of \citet{Saez/Zucman:2014}---the most recent year these data cover---together with our low and high estimates of $\s_k$ as reported in Table \ref{sigmasTab}, then this generates low and high values for the rank-based reversion rates.\footnote{Because this procedure produces two sets of one million different $\a_k$ values, we cannot directly report these estimates in the paper.} These reversion rates generate a perfect match of a stable 2012 U.S. wealth distribution.

As the wealth shares data of \citet{Saez/Zucman:2014} demonstrate, however, stability of the 2012 U.S. distribution of wealth is unlikely. Indeed, a stable distribution is one in which wealth shares are not trending up or down over time, but these data show that the share of total U.S. wealth held by the top 0.01\% and 0.01-0.1\% of households has been steadily rising since the mid-1980s. Our methodology offers several ways to address these stability issues. Most importantly, it is possible to estimate the future stable distribution of wealth using the empirical approach of Section \ref{bench}. 

In order to estimate the future stable distribution, we first observe the rate at which various wealth shares are changing in the economy, and then adjust the rank-based relative growth rates $\a_k$ accordingly. For example, if we observe that the share of total wealth held by the top 1\% of households is increasing at a rate of one percent per year, then we must increase by one percent the value of $\a_k$ for all households in the top 1\%. A similar adjustment must be made for all other subsets of households based on their changing shares of total wealth over time, as well. The future stable distribution that the current distribution of wealth is transitioning towards is then determined by the rank-based reversion rates implied by these adjusted relative growth rates $\a_k$. One of the strengths of this empirical estimation strategy is that it depends only on the rate at which top wealth shares are changing over time and does not rely on any assumptions about the underlying causes of these changes.

The logic behind these adjustments to the parameters $\a_k$ is simple. In a stable distribution, the growth rate of wealth of the $k$-th wealthiest household relative to the whole economy is equal to $\a_k$. Because this distribution is stable, the share of wealth held by the $k$-th wealthiest household $\theta_{(k)}$ should be growing by zero percent per year. If we instead observe that $\theta_{(k)}$ is growing by one percent, then this implies that our estimate of $\a_k$ is one percent too low.\footnote{Of course, a more direct approach is to directly measure the relative growth rates $\a_k$ empirically using panel data. Unfortunately, the lack of a comprehensive panel data set for wealth holdings rules this out.} Indeed, if our estimates of $\a_k$ were correct, then the distribution of wealth would be stable, so any observed instability implies that these estimates must be adjusted.

In order to estimate the future stable distribution of wealth in the U.S., then, we shall use our estimates of mean reversion $-\a_k$ for a stable 2012 U.S. wealth distribution and then adjust these estimates to account for different transitioning wealth shares scenarios. We consider four such scenarios. In the first, we simply assume that the U.S. wealth distribution was stable in 2012. Although this stable scenario is unrealistic given the observed changes in top wealth shares over the past few decades, as discussed above, it is nevertheless a useful baseline case to consider.

In the second scenario, we assume that the share of wealth held by the top 0.01\% of households is increasing by 1\% per year while the share of total wealth held by the bottom 90\% of households is decreasing by 0.5\% per year. No other subset of households is altering its share of total wealth in this scenario. The third scenario posits that the shares of total wealth held by the top 0.01\% and 0.01-0.1\% of households are increasing by 1.5\% and 0.5\% per year, respectively, while the share of total wealth held by the bottom 90\% of households is decreasing by 1\% per year. Finally, in the fourth scenario, we assume that the shares of total wealth held by the top 0.01\% and 0.01-0.1\% of households are increasing by 3\% and 1\% per year, respectively, while the share of total wealth held by the bottom 90\% of households is decreasing by 1.5\% per year.


This process of adjusting the rank-based relative growth rates $\a_k$ to determine the adjusted reversion rates and estimate the future stable distribution of wealth in the U.S. can be applied to any observed changes in wealth shares over time. The four scenarios we consider are intended to capture a range of possible future changes in wealth shares that are of a smaller magnitude than those changes observed in the past few decades, according to the data of \citet{Saez/Zucman:2014}.\footnote{The changes in top shares over this same period according to the SCF are of a smaller magnitude than what is reported by \citet{Saez/Zucman:2014}. Depending upon how the SCF data is adjusted, these data are consistent with an underlying trend somewhere between Scenarios 2 and 4.} Indeed, the magnitude of the changes observed in wealth shares over the past 30 years according to these data is difficult to reconcile with any stable distribution of wealth, a point we discuss in more detail below.

The predicted future stable distributions of wealth for the four scenarios are reported in Tables \ref{futureSharesTab1}-\ref{futureSharesTab4}. These future distributions vary across the low and high estimates of the volatilities $\s_k$, although the tables show that the range of possible outcomes in between these two estimates is in most cases fairly narrow. The four scenarios taken together demonstrate that even a small upward trend in the top wealth shares of today's U.S. economy may imply a substantial increase in wealth concentration once the distribution stabilizes in the future. Indeed, Scenario 3, in which the shares of wealth held by the top 0.01\% and 0.01-0.1\% of households are increasing by a modest 1.5\% and 0.5\% per year, respectively, implies a future stable U.S. distribution of wealth with unprecedented levels of wealth concentration for both the low and high estimates of $\s_k$. Scenario 4 involves an even faster upward trend in inequality than Scenario 3, so that according to Theorem \ref{wealthDistThm2}, this scenario implies that the distribution separates into divergent subpopulations, with the top 0.01\% of households eventually holding all wealth.

\begin{figure}[p]
\begin{center}
\vspace{-5pt}
\hspace{-15pt}\scalebox{.63}{ {\includegraphics{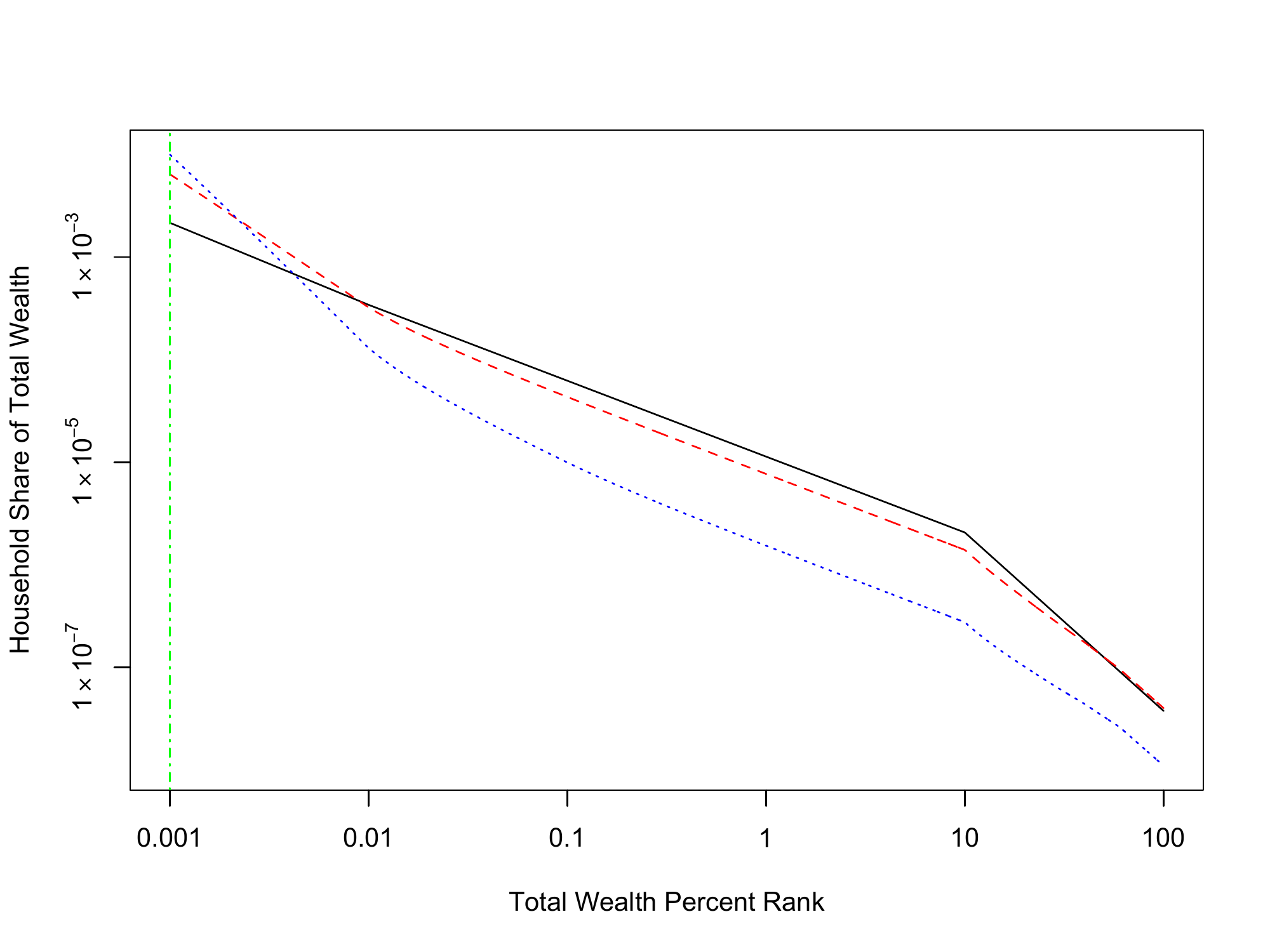}}}
\end{center}
\vspace{-24pt} \caption{Household wealth shares for high estimates of the volatilities $\s_k$ under Scenarios 1 (solid black line), 2 (dashed red line), 3 (dotted blue line), and 4 (vertical dot-dashed green line).}
\label{wealthScenariosFig}
\end{figure}

Figure \ref{wealthScenariosFig} jointly plots the distribution of wealth for all four scenarios using the high estimates of the volatilities $\s_k$. This figure shows just how sensitive the future distribution of wealth is to upward trends in today's top wealth shares. Even though there is a modest difference between the outcomes of Scenarios 1 and 2, once the rate at which the share of wealth held by the top 0.01\% of households is assumed to increase by more than 1\% per year as in Scenarios 3 and 4, the future distribution of wealth transforms dramatically.

These hypothetical estimates of the future U.S. distribution of wealth are not intended as precise quantitative predictions about the future. The future is uncertain and subject to unpredictable changes in those factors that affect the distribution of wealth, such as policy, institutions, technology, and globalization. These factors will change in the future, and those changes will have an effect on the reversion rates $-\a_k$ and hence on the stable distribution of wealth. Forecasting changes in these factors is well beyond the scope of this paper. Instead, the estimates of the future stable U.S. distribution of wealth in Tables \ref{futureSharesTab1}-\ref{futureSharesTab4} are intended to describe the trajectory of inequality in the absence of any future changes in the economic environment. These estimates describe where the U.S. distribution of wealth is currently headed, not whether it will actually get there before something else changes its trajectory.

The absolute concentration of wealth that occurs in Scenario 4 is part of an important set of outcomes in which there is no stable distribution of wealth. This occurs if the share of wealth held by the top 0.01\% of households is increasing rapidly enough, since such rapid increases imply that the adjusted estimates of the rank-based relative growth rates $\a_k$ for many households in the top 0.01\% (those households with $k \leq 100$) are positive. Too many values of $\a_k$ greater than zero violates the stability condition from Theorem \ref{wealthDistThm}, which states that $\a_1 + \cdots + \a_k < 0$, for all $k = 1, \ldots, N - 1$.

In these unstable scenarios, we know from Theorem \ref{wealthDistThm2} that some subset of households is diverging from the rest of the population and forming a separate stable distribution that will eventually hold all wealth. This subset of households eventually separates permanently from the rest of the population so that there is no more movement into or out of this top group. In the case of Scenario 4, this divergent subpopulation consists of the top 0.01\% of households, since most of the adjusted values of the parameters $\a_k$ are positive for this group and negative for the rest of the population.

\begin{figure}[p]
\begin{center}
\vspace{-30pt}
\hspace{-15pt}\scalebox{.63}{ {\includegraphics{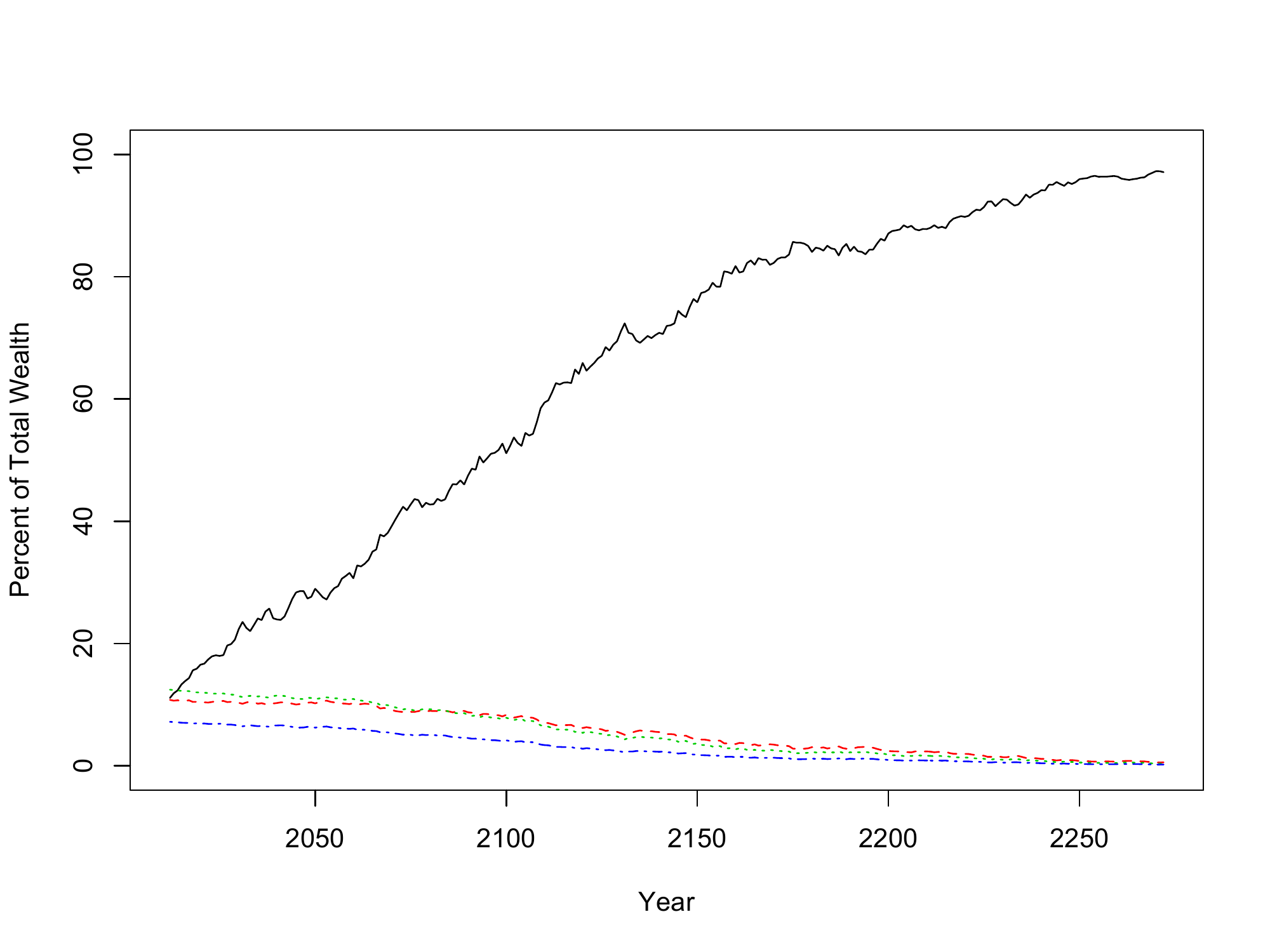}}}
\end{center}
\vspace{-24pt} \caption{The shares of total wealth held by the top 0.01\% (solid black line), 0.01-0.1\% (dashed red line), 0.1-0.5\% (dotted green line), and 0.5-1\% (dot-dashed blue line) of households over time for high estimates of the volatilities $\s_k$ under Scenario 4.}
\label{divergeFig1}
\end{figure}

\begin{figure}[p]
\begin{center}
\vspace{-5pt}
\hspace{-15pt}\scalebox{.63}{ {\includegraphics{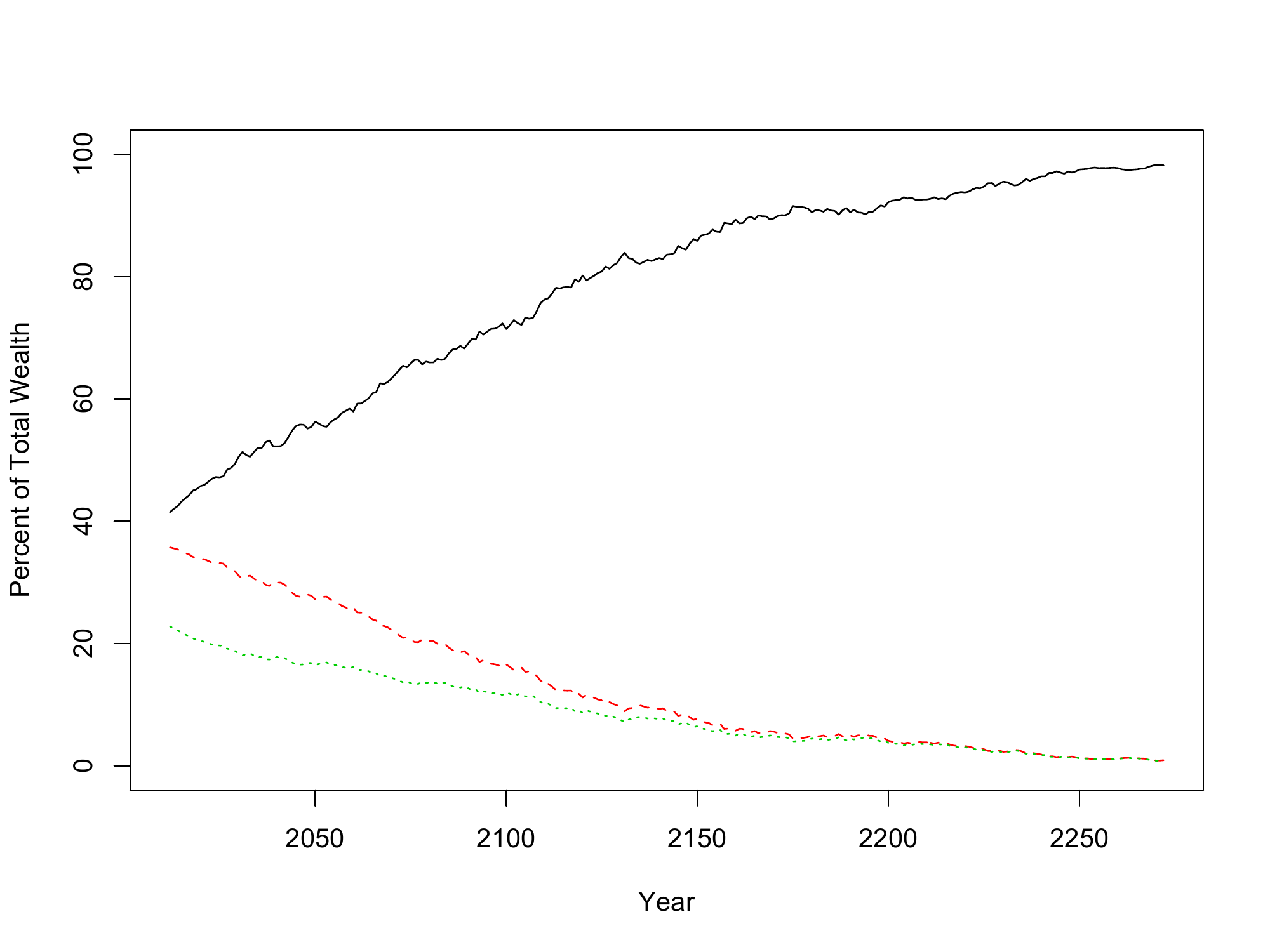}}}
\end{center}
\vspace{-24pt} \caption{The shares of total wealth held by the top 1\% (solid black line), 1-10\% (dashed red line), and bottom 90\% (dotted green line) of households over time for high estimates of the volatilities $\s_k$ under Scenario 4.}
\label{divergeFig2}
\end{figure}

The dynamics by which the U.S. distribution of wealth splits in this scenario are illustrated in Figures \ref{divergeFig1}-\ref{divergeFig2}. Given initial wealth shares equal to those reported by \citet{Saez/Zucman:2014} for the U.S. in 2012, Figure \ref{divergeFig1} shows the simulated evolution over time of the shares of total wealth held by different groups within the top 1\% of households in the economy. While the overall trend of the figure is unmistakable, it is also clear that the divergence of Scenario 4 is uneven, with the share of wealth held by the top 0.01\% of households at times decreasing for several straight years. Over time, however, the top 0.01\% of households gradually but steadily increase their share of total wealth, holding more than 40\% of wealth by 2100 and over 80\% of wealth by 2200. Figure \ref{divergeFig2} shows this same divergent scenario for the top 1\% of households together with the remaining 99\% of households in the economy.




In the real world, of course, it is difficult to imagine that the distribution of wealth would ever truly separate and diverge as in Scenario 4. Long before this occurred, we would expect that some aspect of the economic environment would change. Interestingly, however, it is difficult to reject such a divergent trajectory for the current U.S. distribution of wealth. According to the wealth shares data of \citet{Saez/Zucman:2014}, the share of wealth held by the top 0.01\% of households in the U.S. has increased by an average of more than 3.5\% per year since 2000 and more than 4\% per year since 1980. Changes of this magnitude imply that the adjusted estimates of the rank-based relative growth rates $\a_k$ are positive for the top 0.01\% of households using both the low and high estimates of the volatilities $\s_k$. It might be, then, that according to the wealth shares data of \citet{Saez/Zucman:2014}, the U.S. distribution of wealth is currently on a temporarily divergent trajectory in which a tiny minority of wealthy households indefinitely increases its share of total wealth.


Ultimately, these difficult questions about the current trajectory of the U.S. distribution of wealth and its future direction cannot yet be answered definitively. The exact rate at which top U.S. wealth shares are currently growing is uncertain and varies across different data sets, as can be seen by comparing wealth shares estimates using the income-capitalization method to unadjusted or adjusted estimates from the \emph{Survey of Consumer Finances} \citep{Saez/Zucman:2014}. Only as more detailed and high quality data become available will we be able to provide answers with greater confidence. While this paper does provide preliminary answers and estimates, a more important contribution is to introduce a flexible empirical methodology with which to address these questions.

\subsection{Estimating the Effect of a Progressive Capital Tax}
In principle, the effects of any tax policy on the distribution of wealth can be approximated using our empirical approach. All that is necessary are estimates of the effects of the tax policy on the rank-based reversion rates $-\a_k$ and volatilities $\s_k$. After all, these two factors alone entirely determine the distribution of wealth. In many cases, however, obtaining reliable estimates of the impact of a tax on different households' reversion rates and volatilities of wealth is difficult. One important exception is the case of progressive capital taxes.

The empirical approach of Section \ref{bench} is uniquely suited to estimating the distributional impact of progressive capital taxes since such taxes are designed to have more predictable effects on the growth rates of wealth for different households in the wealth distribution. For simplicity, we shall assume that a capital tax rate of 1\% on some subset of households in the economy reduces the growth rate of wealth for those households by 1\% (and hence also raises their reversion rates by 1\%). Of course, this assumption does not directly take into account the equilibrium effects that such a tax may have on the saving behavior of households and the possibility that households may be able to successfully evade the tax. We choose this simplification because it represents a natural and useful benchmark case in which the reduced savings of households in response to a capital tax (which magnifies the effect of the tax) are exactly balanced by households' ability to evade the tax (which diminishes the effect of the tax). It is important to emphasize that alternative scenarios for the effect of a progressive capital tax are easily evaluated using our empirical approach by simply adjusting the tax's impact on the growth rates of household wealth accordingly.\footnote{One major challenge of estimating the equilibrium effects of a progressive capital tax involves solving the portfolio optimization problem facing households in such an environment. Any further progress towards this end would yield information about how such a tax might alter the growth rates of wealth for different ranked households in equilibrium. This information could then be incorporated into this paper's nonparametric approach to generate general equilibrium estimates of the distributional effect of progressive capital taxes.} Indeed, it is only for brevity that we do not consider such alternative scenarios in this paper.


Ever since \citet{Piketty:2014} proposed a progressive capital tax in response to increasing income and wealth inequality, much of the debate surrounding this policy has centered on how such taxes are likely to increase government revenues or distort economic outcomes rather than how they might affect the distribution of wealth. One of the contributions of this paper is to address this latter issue and provide estimates of the distributional effects of progressive capital taxes on the U.S. economy. These are purely empirical estimates that do not rely on any assumptions about the underlying causes of inequality, a point that is emphasized in the presentation of our model in Section \ref{bench}. 





We analyze a simple progressive capital tax similar to the policy proposed by \citet{Piketty:2014}. Our version of this tax sets the capital tax rate for the top 0.5\% of households in the economy equal to 2\% and the rate for the top 0.5-1\% of households equal to 1\%, while the remaining 99\% of households are assumed to neither pay nor receive any tax or subsidy. As a consequence, none of the revenue generated by the government from this progressive capital tax is redistributed to less wealthy households.\footnote{It is straightforward to consider the distributional effects of such redistribution. However, since we find that these effects are quite small in our model of the U.S wealth distribution, we focus solely on the simple case of a tax without redistribution in this section.} Based on the U.S. wealth distribution data for 2012, this progressive capital tax corresponds to a 2\% rate for those households with total wealth greater than roughly \$6 million and a 1\% rate for those households with total wealth between \$4 and \$6 million.\footnote{The basic progressive capital tax proposed by \citet{Piketty:2014} for Europe involves a 2\% rate for those households with total wealth greater than \euro 5 million, a \%1 rate for those households with total wealth between \euro 1 and \euro 5 million, and no tax for the remaining households.} In terms of the parameters, we assume that this capital tax reduces the rank-based relative growth rates $\a_k$ of those households that are taxed (the top 1\%) by the rate at which they are taxed.

\begin{figure}[p]
\begin{center}
\vspace{-30pt}
\hspace{-15pt}\scalebox{.63}{ {\includegraphics{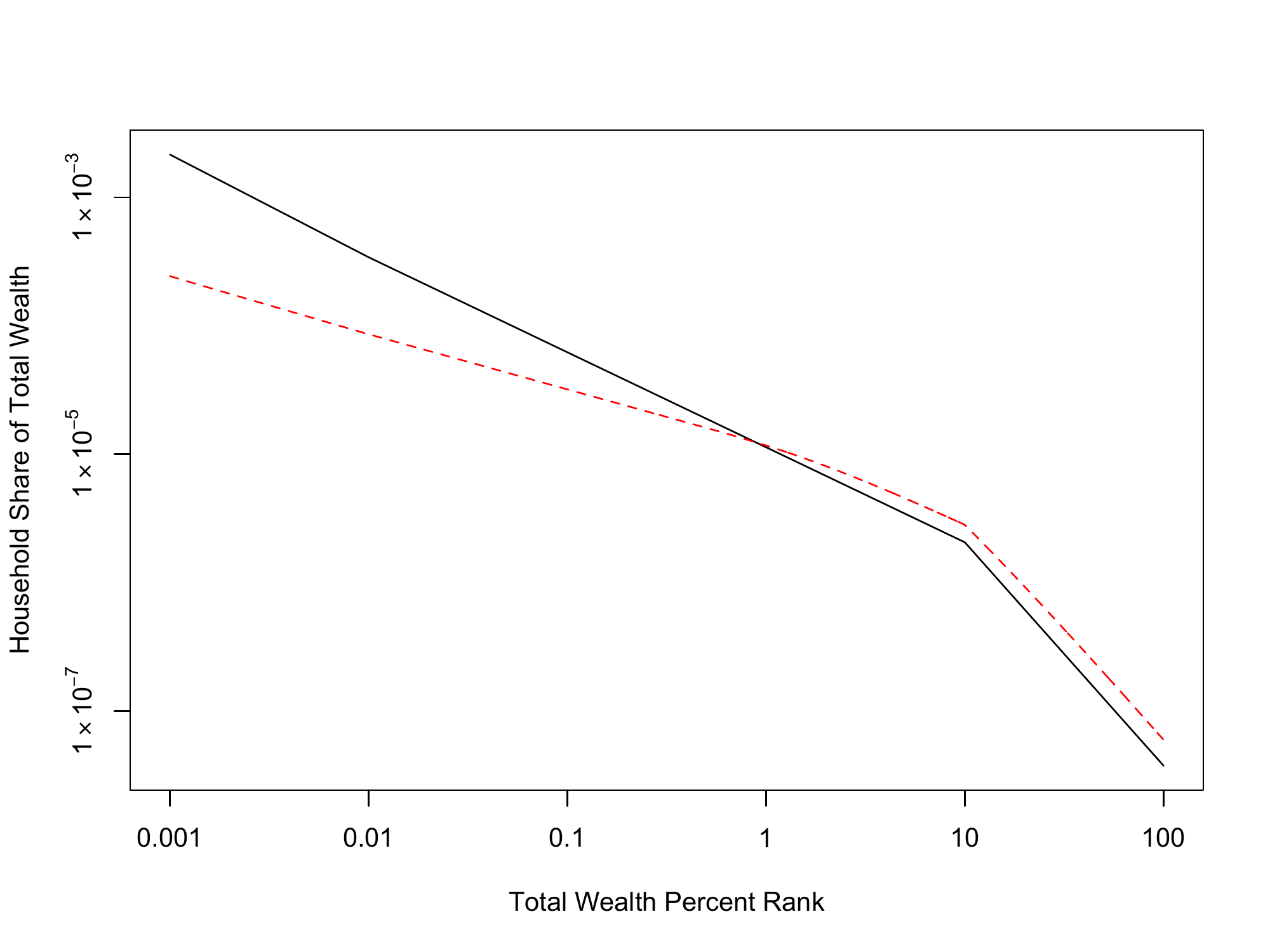}}}
\end{center}
\vspace{-24pt} \caption{Household wealth shares with (dashed red line) and without (solid black line) a 1-2\% progressive capital tax on the top 1\% of households for high estimates of the volatilities $\s_k$ under Scenario 1.}
\label{taxFig1}
\end{figure}

\begin{figure}[p]
\begin{center}
\vspace{-5pt}
\hspace{-15pt}\scalebox{.63}{ {\includegraphics{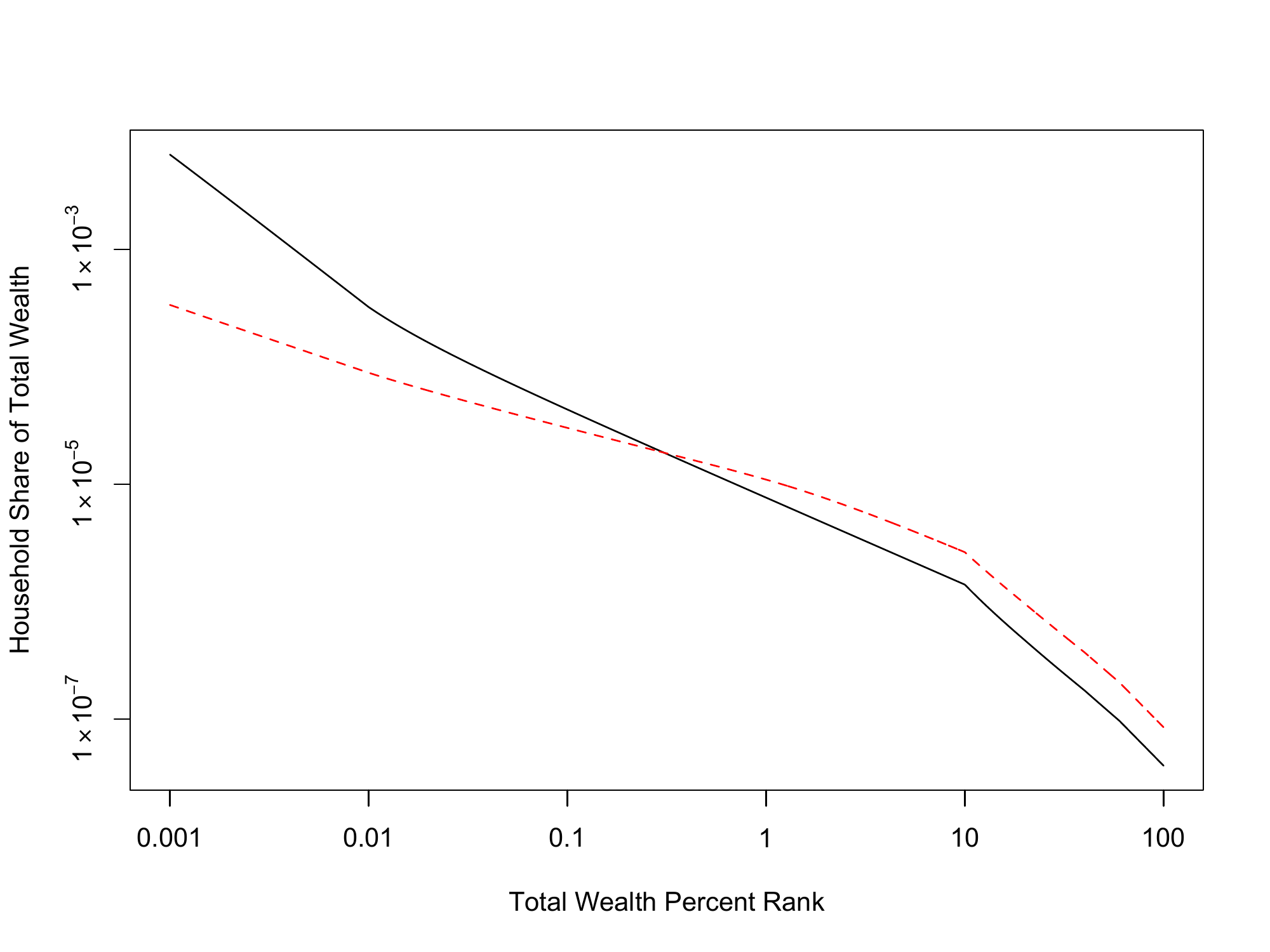}}}
\end{center}
\vspace{-24pt} \caption{Household wealth shares with (dashed red line) and without (solid black line) a 1-2\% progressive capital tax on the top 1\% of households for high estimates of the volatilities $\s_k$ under Scenario 2.}
\label{taxFig2}
\end{figure}

\begin{figure}[p]
\begin{center}
\vspace{-30pt}
\hspace{-15pt}\scalebox{.63}{ {\includegraphics{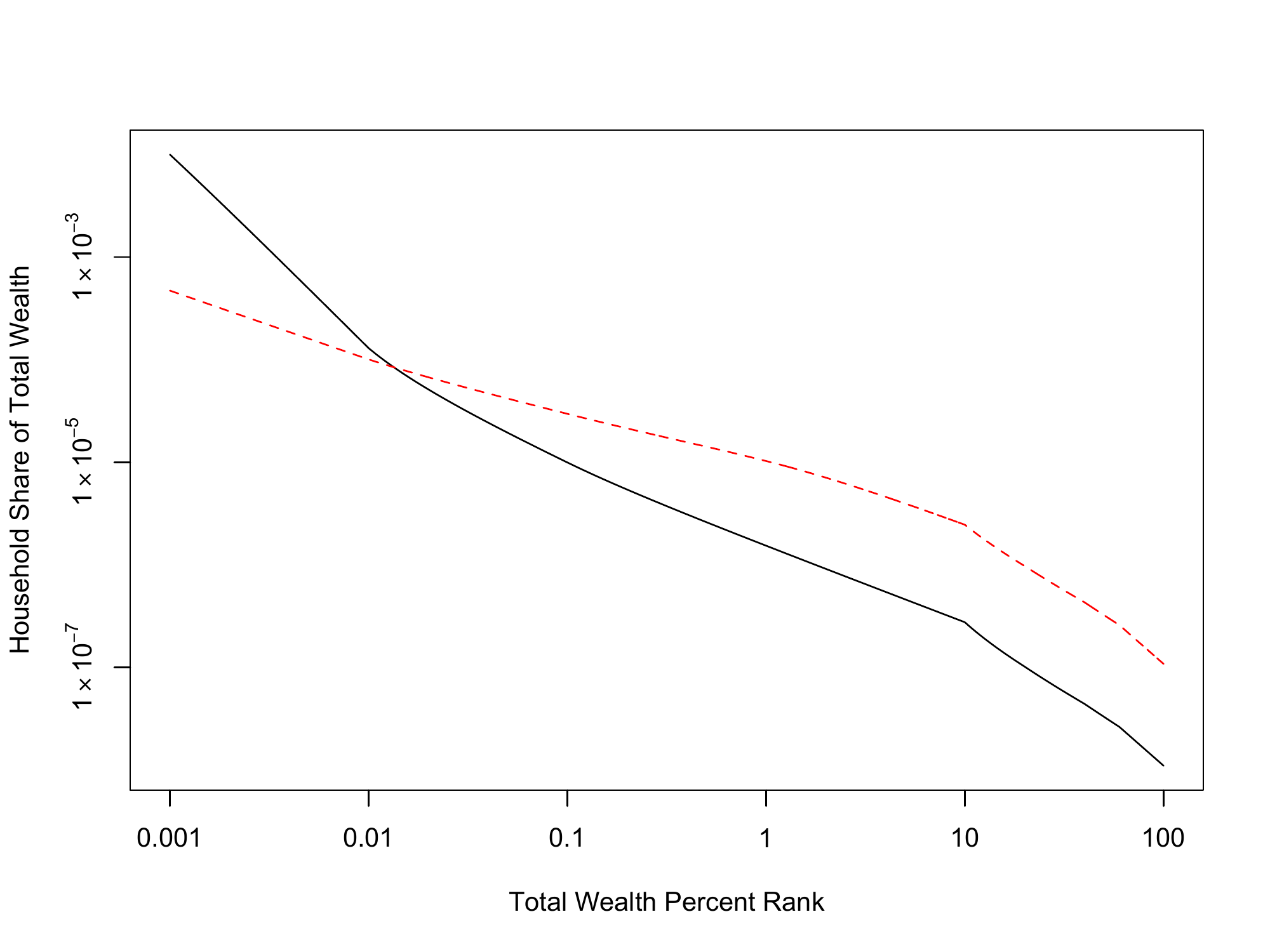}}}
\end{center}
\vspace{-24pt} \caption{Household wealth shares with (dashed red line) and without (solid black line) a 1-2\% progressive capital tax on the top 1\% of households for high estimates of the volatilities $\s_k$ under Scenario 3.}
\label{taxFig3}
\end{figure}

\begin{figure}[p]
\begin{center}
\vspace{-5pt}
\hspace{-15pt}\scalebox{.63}{ {\includegraphics{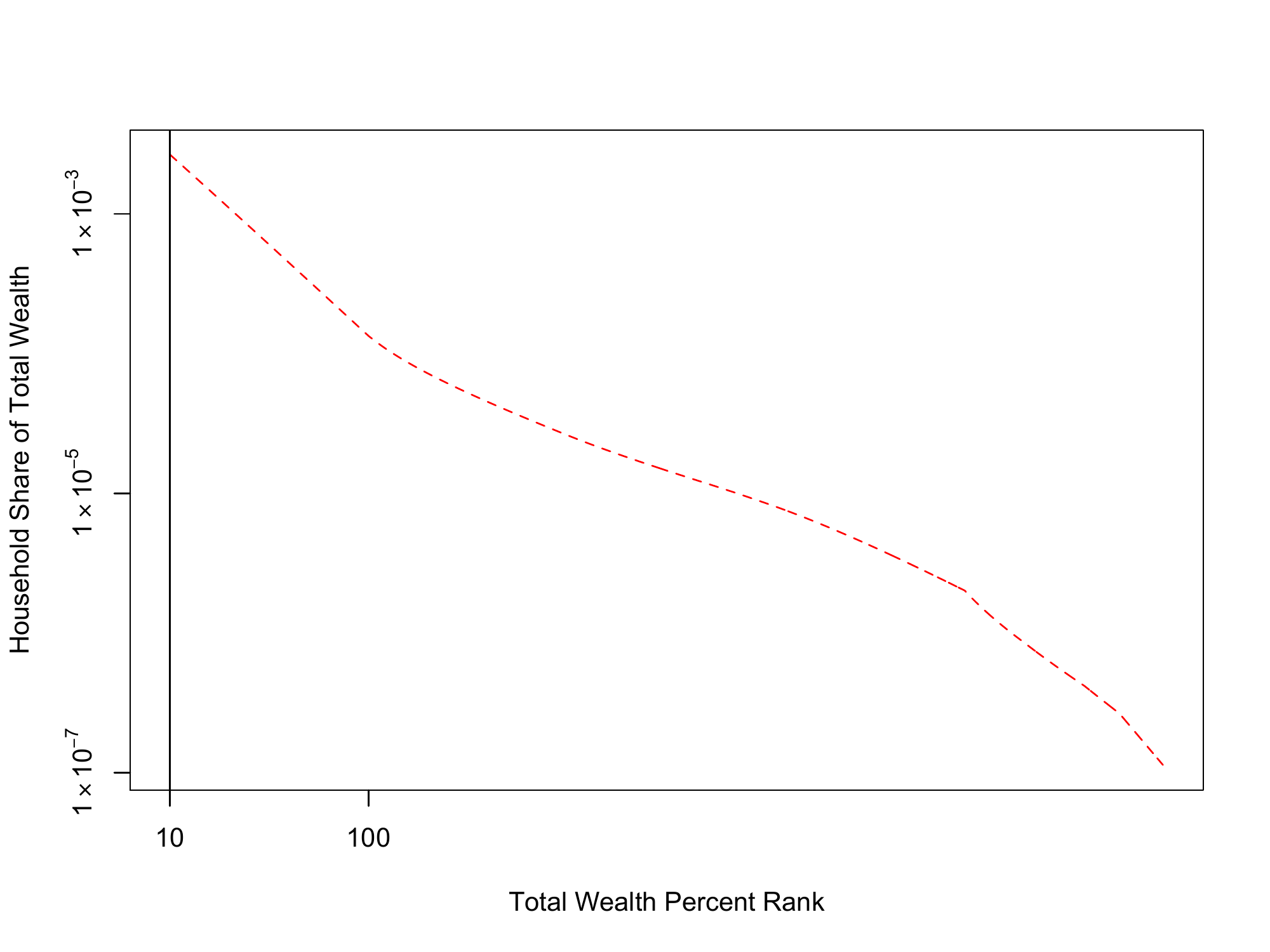}}}
\end{center}
\vspace{-24pt} \caption{Household wealth shares with (dashed red line) and without (vertical solid black line) a 1-2\% progressive capital tax on the top 1\% of households for high estimates of the volatilities $\s_k$ under Scenario 4.}
\label{taxFig4}
\end{figure}

In order to examine a wide range of potential scenarios, we consider the stable distribution of wealth in the presence of this progressive capital tax for Scenarios 1-4 from Tables \ref{futureSharesTab1}-\ref{futureSharesTab4} using the low and high estimates of the volatilities $\s_k$. These results are shown in Tables \ref{taxSharesTab1}-\ref{taxSharesTab4}. They are also shown graphically in Figures \ref{taxFig1}-\ref{taxFig4}, which plot the distribution of wealth both with and without a progressive capital tax for all four scenarios using the high estimates of the volatilities $\s_k$.

In every scenario, the tables and figures show that a simple progressive capital tax imposed on just 1\% of households in the economy substantially reshapes the distribution of wealth and reduces inequality. In the case of Scenarios 1 and 2, for example, the after-tax stable distribution of wealth is similar to the distribution observed in the U.S. in the 1970s, according to the historical wealth shares data of \citet{Saez/Zucman:2014}. The fact that this period is one of the most egalitarian in the U.S. in the last century highlights just how significant this reduction in inequality is. Even for Scenarios 3 and 4, in which the distribution of wealth is either highly or fully concentrated at the top in the absence of a progressive capital tax, in all cases the tax reduces inequality to levels lower than those observed in the U.S. in 2012. This large effect is seen clearly in Figure \ref{taxFig4}. Because there is no stable distribution using the high estimates of $\s_k$ in the case of Scenario 4, this figure represents the distribution of wealth without a progressive capital tax by a vertical line indicating that the top 0.01\% of households hold all the wealth in the economy. Only with the tax in place does a stable distribution exist, as shown by the dashed red line in the figure.

How is it that a 1-2\% progressive capital tax levied on only 1\% of households can reduce inequality so substantially? One might expect that such a large reduction in inequality requires that a larger subset of households be taxed. However, because the top 1\% of households that pay the tax hold between 40-100\% of total wealth depending upon the scenario, a large fraction of the economy's total wealth is in fact affected by this progressive capital tax. As the model demonstrates, this large fraction of total wealth is sufficient for the tax to significantly reshape the distribution of wealth.

Finally, we stress that this result is not a statement about total welfare and not an endorsement of a progressive capital tax. As discussed in the introduction, our statistical model only generates empirical estimates of the distributional effects of taxes and other policies, it does not measure any distortions or costs associated with such policies. Our results about the effects of progressive capital taxes on inequality are intended only to add to our knowledge of the overall effects of such a policy.

\vskip 70pt

\section{Conclusion} \label{conclusion}

In this paper, we have developed a statistical model of inequality in which heterogeneous households are subject to both aggregate and idiosyncratic fluctuations in their wealth holdings. The model imposes few restrictions and no parametric structure on these fluctuations of household wealth. In this setting, we apply new techniques to obtain a closed-form household-by-household characterization of the stable distribution of wealth. According to this characterization, the distribution of wealth is shaped entirely by two factors---the reversion rates and idiosyncratic volatilities of wealth for different ranked households. The simplicity and generality of this result suggests that to understand the effect of factors such as policy, institutions, technology, or globalization on inequality, it is necessary only to understand the effect of these factors on the reversion rates and idiosyncratic volatilities of wealth. As a consequence, more detailed empirical work focused on accurately measuring the reversion rates and volatilities of wealth as well as any variation in these two factors over time and across different regions is likely to yield substantial new insights.

Our statistical model can exactly match any empirical distribution, and we use the wealth shares data of \citet{Saez/Zucman:2014} to construct such a match for the 2012 U.S. distribution of wealth. One of the challenges in analyzing this distribution is that these wealth shares data show a clear upward trend in top shares over the past three decades. These upward trends imply that any analysis that relies on a stable or steady-state distribution of wealth is flawed. This paper introduces a methodology that can address these stability issues. In particular, our approach allows us to estimate the future stable distribution of wealth in the presence of trends in top wealth shares. We present such estimates for several alternative scenarios, but the most likely scenario might be that according to the wealth shares data of \citet{Saez/Zucman:2014}, the U.S. distribution of wealth is on a temporarily unstable trajectory in which it separates into two divergent subpopulations. We also present estimates of the distributional implications of progressive capital taxes. Specifically, we consider a capital tax of 1-2\% levied on 1\% of households similar to the tax proposed by \citet{Piketty:2014}. Although the full effect of this tax depends on the uncertain future stable distribution of wealth, in all cases we find that this tax substantially reduces inequality and reshapes the distribution of wealth.

The statistical model that we develop in this paper is based on a general approach to rank-based systems. Although this approach is well-suited to modeling the distribution of wealth, it is not restricted to only modeling wealth. In fact, only in the case of unstable or i.i.d.-like processes is our general approach clearly not appropriate. This means that there are other areas of economics, such as the distribution of income and the world distribution of output, in which our tractable solution techniques may provide new information.

\pagebreak

\begin{spacing}{1.1}

\appendix
\section{Assumptions and Regularity Conditions} \label{assumptions}

In this appendix, we present the assumptions and regularity conditions that are necessary for the stable wealth distribution characterization in Theorem \ref{wealthDistThm}. As discussed in Section \ref{bench}, these assumptions admit a large class of continuous wealth processes for the households in the economy. The first assumption establishes basic integrability conditions that are common for both continuous semimartingales and It{\^o} processes.

\begin{ass} \label{basicAss}
For all $i = 1, \ldots, N$, the growth rate processes $\m_i$ satisfy
\begin{equation} \label{growthIntegrability}
 \intT|\m_i(t)|\,dt < \infty, \quad \text{$T > 0$,} \as,
\end{equation}
and the volatility processes $\d_i$ satisfy
\begin{align}
 & \intT\left(\d^2_1(t) + \cdots + \d^2_M(t)\right)\,dt < \infty, \quad \text{$T > 0$,} \as, \label{volIntegrability} \\
 & \d^2_1(t) + \cdots + \d^2_M(t) > 0, \quad \text{$t > 0$,} \as \label{volPositive} \\
 & \limt1\left(\d^2_1(t) + \cdots + \d^2_M(t)\right) \log\log t = 0, \label{volBounded} \as,
\end{align}
\end{ass}

Conditions \eqref{growthIntegrability} and \eqref{volIntegrability} are standard in the definition of an It{\^o} process, while condition \eqref{volPositive} ensures that household wealth holdings contain a nonzero random component at all times. Condition \eqref{volBounded} is similar to a boundedness condition in that it ensures that the variance of household wealth holdings does not diverge to infinity too rapidly.

The second assumption underlying our results establishes that no two households' wealth holdings be perfectly correlated over time. In other words, there must always be some idiosyncratic component to household wealth dynamics. Finally, we also assume that no household's wealth holdings relative to the economy shall disappear too rapidly.

\begin{ass} \label{corrAss}
The symmetric matrix $\r(t)$, given by $\r(t) = (\r_{ij}(t))$, where $1 \leq i, j \leq N$, is nonsingular for all $t > 0$, a.s.
\end{ass}

\begin{ass} \label{coherentAss}
For all $i = 1, \ldots, N$, the wealth share processes $\theta_i$ satisfy
\begin{equation}
\limt1\log\theta_i(t) = 0, \as
\end{equation}
\end{ass}

\vskip 50pt

\section{Proofs} \label{proofs}

This appendix presents the proofs of Lemmas \ref{totalWealthLemma} and \ref{localTimeLemma} and Theorems \ref{wealthDistThm} and \ref{wealthDistThm2}.

\begin{proofLemma1}
By definition, $w(t) = w_1(t) + \cdots + w_N(t)$ and for all $i = 1, \ldots, N$, $\theta_i(t) = w_i(t)/w(t)$. This implies that
\begin{equation*}
 dw(t) = \sum_{i=1}^Ndw_i(t) = \sum_{i=1}^N\theta_i(t)w(t)\frac{dw_i(t)}{w_i(t)},
\end{equation*}
from which it follows that
\begin{equation} \label{totalWealthDynamicsProof}
 \frac{dw(t)}{w(t)} = \sum_{i=1}^N\theta_i(t)\frac{dw_i(t)}{w_i(t)}.
\end{equation}
We wish to show that the process satisfying equation \eqref{totalWealthDynamics} also satisfies equation \eqref{totalWealthDynamicsProof}.

If we apply \ito's Lemma to the exponential function, then equation \eqref{totalWealthDynamics} yields
\begin{equation} \label{totalWealthDynamicsProof2}
\begin{aligned}
 dw(t) & = w(t)\m(t)\,dt + \frac{1}{2}w(t)\sum_{i,j=1}^N\theta_i(t)\theta_j(t)\left(\sum_{z=1}^M\d_{iz}(t)\d_{jz}(t)\right)\,dt \\
  & \qquad + w(t)\sum_{i=1}^N\sum_{z=1}^M\theta_i(t)\d_{iz}(t)\,dB_z(t),
\end{aligned}
\end{equation}
a.s., where $\m(t)$ is given by equation \eqref{mu}. Using the definition of $\r_{ij}(t)$ from equation \eqref{rhoIJ}, we can simplify equation \eqref{totalWealthDynamicsProof} and write
\begin{equation} \label{totalWealthDynamicsProof3}
\frac{dw(t)}{w(t)} = \left(\m(t) + \frac{1}{2}\sum_{i,j=1}^N\theta_i(t)\theta_j(t)\r_{ij}(t)\right)\,dt + \sum_{i=1}^N\sum_{z=1}^M\theta_i(t)\d_{iz}(t)\,dB_z(t).
\end{equation}
Similarly, the definition of $\m(t)$ from equation \eqref{mu} allows us to further simplify equation \eqref{totalWealthDynamicsProof3} and write
\begin{align}
\frac{dw(t)}{w(t)} & = \left(\sum_{i=1}^N\theta_i(t)\m_i(t) + \frac{1}{2}\sum_{i=1}^N\theta_i(t)\r_{ii}(t)\right)\,dt +\sum_{i=1}^N\sum_{z=1}^M\theta_i(t)\d_{iz}(t)\,dB_z(t) \notag \\
& = \sum_{i=1}^N\theta_i(t)\left(\m_i(t) + \frac{1}{2}\r_{ii}(t)\right)\,dt + \sum_{i=1}^N\sum_{z=1}^M\theta_i(t)\d_{iz}(t)\,dB_z(t).\label{totalWealthDynamicsProof4}
\end{align}

If we again apply \ito's Lemma to the exponential function, then equation \eqref{wealthDynamics} yields, a.s., for all $i = 1, \ldots, N$,
\begin{align}
 dw_i(t) & = w_i(t)\left(\m_i(t) + \frac{1}{2}\sum_{z=1}^M\d^2_{iz}(t)\right)\,dt + w_i(t)\sum_{z=1}^M\d_{iz}(t)\,dB_z(t) \notag \\
 & =  w_i(t)\left(\m_i(t) + \frac{1}{2}\r_{ii}(t)\right)\,dt + w_i(t)\sum_{z=1}^M\d_{iz}(t)\,dB_z(t). \label{wealthDynamicsProof}
\end{align}
Substituting equation \eqref{wealthDynamicsProof} into equation \eqref{totalWealthDynamicsProof4} then yields
\begin{equation*}
 \frac{dw(t)}{w(t)} = \sum_{i=1}^N\theta_i(t)\frac{dw_i(t)}{w_i(t)},
\end{equation*}
which completes the proof.
\end{proofLemma1}

\begin{proofLemma2}
The household wealth processes $w_i$ are absolutely continuous in the sense that the random signed measures $\m_i(t)\,dt$ and $\r_{ii}(t)\,dt$ are absolutely continuous with respect to Lebesgue measure. As a consequence, we can apply Lemma 4.1.7 and Proposition 4.1.11 from \citet{Fernholz:2002}, which yields equations \eqref{rankWealthDynamics} and \eqref{rankWealthShareDynamics1}.
\end{proofLemma2}

\begin{proofTheorem1}
This proof follows arguments from Chapter 5 of \citet{Fernholz:2002}. According to equation \eqref{rankWealthShareDynamics2}, for all $k = 1, \ldots, N$,
\begin{equation} \label{rankWealthShareProof1}
\begin{aligned}
 \log\theta_{(k)}(T) & = \intT\left(\m_{p_t(k)}(t) - \m(t)\right)\,dt +  \frac{1}{2}\L_{\log\theta_{(k)} - \log\theta_{(k + 1)}}(T) - \frac{1}{2}\L_{\log\theta_{(k - 1)} - \log\theta_{(k)}}(T) \\
 & \qquad + \sum_{z=1}^M\intT\d_{p_t(k)z}(t)\,dB_z(t) - \sum_{i=1}^N\sum_{z=1}^M\intT\theta_i(t)\d_{iz}(t)\,dB_z(t).
\end{aligned}
\end{equation}
Consider the asymptotic behavior of the process $\log\theta_{(k)}$. Assuming that the limits from equation \eqref{kappa} exist, then according to the definition of $\a_k$ from equation \eqref{alphaK}, the asymptotic behavior of $\log\theta_{(k)}$ satisfies
\begin{equation} \label{rankWealthShareProof2}
\begin{aligned}
 \limT1\log\theta_{(k)}(T) & = \a_k + \frac{1}{2}\k_k - \frac{1}{2}\k_{k-1} + \limT1\sum_{z=1}^M\intT\d_{p_t(k)z}(t)\,dB_z(t) \\
 & \qquad  - \limT1\sum_{i=1}^N\sum_{z=1}^M\intT\theta_i(t)\d_{iz}(t)\,dB_z(t), \as
\end{aligned}
\end{equation}
Assumption \ref{coherentAss} ensures that the term on the left-hand side of equation \eqref{rankWealthShareProof2} is equal to zero, while Assumption \ref{basicAss} ensures that the last two terms of the right-hand side of this equation are equal to zero as well (see Lemma 1.3.2 from \citealp{Fernholz:2002}). If we simplify equation \eqref{rankWealthShareProof2}, then, we have that
\begin{equation} \label{alphaKProof}
 \a_k = \frac{1}{2}\k_{k-1} - \frac{1}{2}\k_k,
\end{equation}
which implies that
\begin{equation} \label{alphaKappa}
 \a_k - \a_{k+1} = \frac{1}{2}\k_{k-1} - \k_k + \frac{1}{2}\k_{k+1},
\end{equation}
for all $k = 1, \ldots, N-1$. Since equation \eqref{alphaKProof} is valid for all $k = 1, \ldots, N$, this establishes a system of equations that we can solve for $\k_k$. Doing this yields the equality
\begin{equation} \label{kappaAlphaProof}
 \k_k = -2(\a_1 + \cdots + \a_k),
\end{equation}
for all $k = 1, \ldots, N$. Note that asymptotic stability ensures that $\a_1 + \cdots + \a_k < 0$ for all $k = 1, \ldots, N$, while the fact that $\a_N = \frac{1}{2}\k_{N-1} = -(\a_1 + \cdots + \a_{N-1})$ ensures that $\a_1 + \cdots + \a_N = 0$. Furthermore, if $\a_1 + \cdots + \a_k > 0$ for some $1 \leq k < N$, then equation \eqref{kappaAlphaProof} generates a contradiction since $\k_k \geq 0$ by definition. In this case, it must be that Assumption \ref{coherentAss} is violated and $\limT1\log\theta_{(k)}(T) \neq 0$ for some $1 \leq k \leq N$. This case is examined in detail in Theorem \ref{wealthDistThm2}.

The last term on the right-hand side of equation \eqref{rankWealthShareDynamics3} is an absolutely continuous martingale, and hence can be represented as a stochastic integral with respect to Brownian motion $B(t)$.\footnote{This is a standard result for continuous-time stochastic processes \citep{Karatzas/Shreve:1991,Nielsen:1999}.} This fact, together with equation \eqref{alphaKappa} and the definitions of $\a_k$ and $\s_k$ from equations \eqref{alphaK}-\eqref{sigmaK}, motivates our use of the stable version of the process $\log\theta_{(k)} - \log\theta_{(k+1)}$. Recall that, by equation \eqref{stableVersion}, this stable version is given by
\begin{equation} \label{rankWealthShareDynamicsProof}
 d\left(\log\hat{\theta}_{(k)}(t) - \log\hat{\theta}_{(k+1)}(t)\right) = -\k_k\,dt + d\L_{\log\hat{\theta}_{(k)} - \log\hat{\theta}_{(k + 1)}}(t) + \s_k\,dB(t),
\end{equation}
for all $k = 1, \ldots, N-1$. According to \citet{Fernholz:2002}, Lemma 5.2.1, for all $k = 1, \ldots, N-1$, the time-averaged limit of this stable version satisfies
\begin{equation} \label{thetaKProof}
 \limT1\intT\left(\log\hat{\theta}_{(k)}(t) - \log\hat{\theta}_{(k+1)}(t)\right)\,dt = \frac{\s^2_k}{2\k_k} = \frac{\s^2_k}{-4(\a_1 + \cdots + \a_k)},
\end{equation}
a.s., where the last equality follows from equation \eqref{kappaAlphaProof}. To the extent that the stable version of $\log\theta_{(k)} - \log\theta_{(k+1)}$ from equation \eqref{rankWealthShareDynamicsProof} approximates the true version of this process from equation \eqref{rankWealthShareDynamics3}, the time-averaged limit of the true process $\log\theta_{(k)} - \log\theta_{(k+1)}$ will be approximated by $-\s^2_k/4(\a_1 + \cdots + \a_k)$, for all $k = 1, \ldots, N-1$.
\end{proofTheorem1}

\begin{proofTheorem2}
Note that the divergent scenario of Theorem \ref{wealthDistThm2} violates Assumption \ref{coherentAss}, which states that no household's share of wealth declines to zero too quickly. In order to prove this theorem, it is necessary to show that the largest subset of households for which Assumption \ref{coherentAss} holds is also the subset of households $m < N$ satisfying $A_m = \max_{1 \leq k \leq N} A_k$ and $A_m > A_l$ for $l \neq m$.

Suppose that the top $n \leq N$ wealthiest households in the economy form the largest subset of households for which Assumption \ref{coherentAss} holds. More precisely, suppose that
\begin{equation} \label{limProof1}
 \limT1\log\theta_{(1)}(T) = \cdots = \limT1\log\theta_{(n)}(T) = 0, \as,
\end{equation}
and
\begin{equation} \label{limProof2}
 0 > \limT1\log\theta_{(n+1)}(T) \geq \cdots \geq \limT1\log\theta_{(N)}(T), \as,
\end{equation}
so that there exists some finite time $t_0$ such that after $t_0$, the top $n$ wealthiest households are never overtaken by the remaining households in the economy again.\footnote{The existence of the limits in equation \eqref{limProof2} is ensured by equation \eqref{rankWealthShareProof2} together with the assumption that the limits from equations \eqref{alphaK}-\eqref{kappa} exist.} Indeed, this follows because equations \eqref{limProof1} and \eqref{limProof2} imply that there exists some $t_0 < \infty$ such that $\log w_{(n)}(t) > \log w_{(n+1)}(t)$, a.s., for all $t \geq t_0$ and that
\begin{equation} \label{limThetasProof}
 \lim_{T\to\infty}\theta_{(1)}(T) + \cdots + \theta_{(n)}(T) = 1, \as
\end{equation}
Without loss of generality, let the wealth holdings of the $n$ wealthiest households in the economy at time $t_0$ be $w_1(t_0), \ldots, w_n(t_0)$.

Consider the wealth holdings of the top $n$ wealthiest households in the economy, which we denote by
\begin{equation*}
 w^n(t) = w_{(1)}(t) + \cdots + w_{(n)}(t) = w_{p_t(1)}(t) + \cdots + w_{p_t(n)}(t).
\end{equation*}
Because Assumption \ref{coherentAss} is valid for the $n$ wealthiest households in the economy at time $t_0$, according to \citet{Fernholz:2002}, Propositions 1.3.1 and 2.1.2, and using Assumption \ref{basicAss}, equation \eqref{volBounded}, it follows that
\begin{equation} \label{limWnProof1}
 \limT{\log w^n(T)} = \limT1\intT\m_i(t)\,dt, \as,
\end{equation}
where $1 \leq i \leq n$. By definition, for all $i = 1, \ldots, n$,
\begin{equation*}
 \m_i(t) = \sum_{k=1}^n\m_{p_t(k)}(t)I_{\{0\}}\left(\theta_i(t) - \theta_{(k)}(t)\right),
\end{equation*}
and hence by equation \eqref{limWnProof1},
\begin{equation} \label{limWnProof2}
 \limT{\log w^n(T)} = \limT1\intT\sum_{k=1}^n\m_{p_t(k)}(t)I_{\{0\}}\left(\theta_i(t) - \theta_{(k)}(t)\right)\,dt, \as
\end{equation}
If the households in the economy are ex-ante symmetric, then all of them must spend equal fractions of time in any given rank \citep{Banner/Fernholz/Karatzas:2005}. As a consequence, for all $k = 1, \ldots, n$,
\begin{equation*}
 \limT1\intT I_{\{0\}}\left(\theta_i(t) - \theta_{(k)}(t)\right)\,dt = \frac{1}{n}, \as,
\end{equation*}
which together with equation \eqref{limWnProof2} implies that
\begin{equation*}
 \limT{\log w^n(T)} = \limT1\intT\frac{1}{n}\sum_{k=1}^n\m_{p_t(k)}(t)\,dt, \as
\end{equation*}
By the definition of $A_n$ from equation \eqref{Am}, it follows that
\begin{equation*}
 \limT{\log w^n(T)} = A_n + \limT1\intT\m(t)\,dt, \as
\end{equation*}
Of course, because $w^n$ converges to $w$ over time, it follows by equation \eqref{limThetasProof} that $A_n = 0$. Intuitively, the relative growth rate of $w^n$ must equal zero since $w^n$ gradually encompasses all wealth in the economy.

Suppose that $m < n$. Because Assumption \ref{coherentAss} is valid for the $n$ wealthiest households in the economy at time $t_0$, we can simply reproduce the proof of Theorem \ref{wealthDistThm} for this top subset of households. However, if $A_m > A_n$, equation \eqref{kappaAlphaProof} generates a contradiction since it implies that
\begin{equation*}
  \k_m = -(\a_1 + \cdots + \a_m) = -mA_m < 0,
\end{equation*}
while $\k_k \geq 0$ for all $1 \leq k \leq N$, by definition from equation \eqref{kappa} (see also the discussion in the proof of Theorem \ref{wealthDistThm}). Thus, we conclude that $m \geq n$.

Suppose that $m > n$, so that by definition
\begin{equation} \label{AmIneqProof}
 \frac{\a_{n+1} + \cdots + \a_m}{m - n} > A_n.
\end{equation}
According to equation \eqref{rankWealthShareProof2} from the proof of Theorem \ref{wealthDistThm}, we have that
\begin{equation} \label{limProof4}
 \limT1\log\theta_{(n+1)}(T) + \cdots + \limT1\log\theta_{(m)}(T) = \a_{n+1} + \cdots + \a_m + \frac{1}{2}\k_m - \frac{1}{2}\k_n, \as
\end{equation}
Of course, by assumption $\k_n = 0$, since after $t_0$ the top $n$ wealthiest households are never overtaken by the remaining households in the economy again (recall the definition of a local time $\L_x$). Furthermore, by equation \eqref{AmIneqProof}, it follows that $\a_{n+1} + \cdots + \a_m > 0$ and hence that the right-hand side of equation \eqref{limProof4} is greater than zero ($\k_m \geq 0$ as well). This is a contradiction, however, since we assumed in equation \eqref{limProof2} above that the left-hand side of equation \eqref{limProof4} is less than zero. Thus, we conclude that $m = n$ and that the largest subset of households for which Assumption \ref{coherentAss} holds is also the subset of households $m < N$ satisfying $A_m = \max_{1 \leq k \leq N} A_k$ and $A_m > A_l$ for $l \neq m$.

Having proved the separation and divergence of the top subset of households $w_{(1)}, \ldots, w_{(m)}$, all that remains is to prove that this subset forms a stable distribution. This follows from Theorem \ref{wealthDistThm} and the fact that $A_m > A_l$ for $l \neq m$, since this condition ensures that the relative growth rates for this top subset of households satisfies the stability condition of Theorem \ref{wealthDistThm}.
\end{proofTheorem2}


\end{spacing}

\pagebreak

\begin{spacing}{1.2}

\bibliographystyle{chicago}
\bibliography{econ}

\begin{thebibliography}{}

\bibitem[\protect\citeauthoryear{Aiyagari}{Aiyagari}{1994}]{Aiyagari:1994}
Aiyagari, S.~R. (1994, August).
\newblock Uninsured idiosyncratic risk and aggregate saving.
\newblock {\em Quarterly Journal of Economics\/}~{\em 109\/}(3), 659--684.

\bibitem[\protect\citeauthoryear{Altonji, {Smith Jr.}, and Vidangos}{Altonji
  et~al.}{2013}]{Altonji/Smith/Vidangos:2013}
Altonji, J.~G., A.~A. {Smith Jr.}, and I.~Vidangos (2013, July).
\newblock Modeling earnings dynamics.
\newblock {\em Econometrica\/}~{\em 81\/}(4), 1395--1454.

\bibitem[\protect\citeauthoryear{Angeletos}{Angeletos}{2007}]{Angeletos:2007}
Angeletos, G.-M. (2007, January).
\newblock Uninsured idiosyncratic investment risk and aggregate saving.
\newblock {\em Review of Economic Dynamics\/}~{\em 10\/}(1), 1--30.

\bibitem[\protect\citeauthoryear{Angeletos and Calvet}{Angeletos and
  Calvet}{2006}]{Angeletos/Calvet:2006}
Angeletos, G.-M. and L.-E. Calvet (2006, September).
\newblock Idiosyncratic production risk, growth, and the business cycle.
\newblock {\em Journal of Monetary Economics\/}~{\em 53\/}(6), 1095--1115.

\bibitem[\protect\citeauthoryear{Atkinson, Piketty, and Saez}{Atkinson
  et~al.}{2011}]{Atkinson/Piketty/Saez:2011}
Atkinson, A.~B., T.~Piketty, and E.~Saez (2011, March).
\newblock Top incomes in the long run of history.
\newblock {\em Journal of Economic Literature\/}~{\em 49\/}(1), 3--71.

\bibitem[\protect\citeauthoryear{Banner, Fernholz, and Karatzas}{Banner
  et~al.}{2005}]{Banner/Fernholz/Karatzas:2005}
Banner, A., R.~Fernholz, and I.~Karatzas (2005).
\newblock Atlas models of equity markets.
\newblock {\em Annals of Applied Probability\/}~{\em 15\/}(4), 2296--2330.

\bibitem[\protect\citeauthoryear{Benhabib, Bisin, and Zhu}{Benhabib
  et~al.}{2011}]{Benhabib/Bisin/Zhu:2011}
Benhabib, J., A.~Bisin, and S.~Zhu (2011, January).
\newblock The distribution of wealth and fiscal policy in economies with
  infinitely lived agents.
\newblock {\em Econometrica\/}~{\em 79\/}(1), 123--157.

\bibitem[\protect\citeauthoryear{Benhabib, Bisin, and Zhu}{Benhabib
  et~al.}{2014}]{Benhabib/Bisin/Zhu:2014}
Benhabib, J., A.~Bisin, and S.~Zhu (2014).
\newblock The distribution of wealth in the blanchard-yaari model.
\newblock Forthcoming in \emph{Macroeconomic Dynamics}.

\bibitem[\protect\citeauthoryear{Bonhomme and Robin}{Bonhomme and
  Robin}{2010}]{Bonhomme/Robin:2010}
Bonhomme, S. and J.-M. Robin (2010, April).
\newblock Generalized non-parametric deconvolution with an application to
  earnings dynamics.
\newblock {\em Review of Economic Studies\/}~{\em 77\/}(2), 491--533.

\bibitem[\protect\citeauthoryear{Browning, Ejrn{\ae}s, and Alvarez}{Browning
  et~al.}{2010}]{Browning/Ejrnaes/Alvarez:2010}
Browning, M., M.~Ejrn{\ae}s, and J.~Alvarez (2010, October).
\newblock Modelling income processes with lots of heterogeneity.
\newblock {\em Review of Economic Studies\/}~{\em 77\/}(4), 1353--1381.

\bibitem[\protect\citeauthoryear{Cagetti and {De Nardi}}{Cagetti and {De
  Nardi}}{2008}]{Cagetti/DeNardi:2008}
Cagetti, M. and M.~{De Nardi} (2008, September).
\newblock Wealth inequality: Data and models.
\newblock {\em Macroeconomic Dynamics\/}~{\em 12\/}(S2), 285--313.

\bibitem[\protect\citeauthoryear{Case and Shiller}{Case and
  Shiller}{1989}]{Case/Shiller:1989}
Case, K.~E. and R.~J. Shiller (1989, March).
\newblock The efficiency of the market for single-family homes.
\newblock {\em American Economic Review\/}~{\em 79\/}(1), 125--137.

\bibitem[\protect\citeauthoryear{Casta{\~n}eda, D\'{i}az-Gim\'{e}nez, and
  R\'{i}os-Rull}{Casta{\~n}eda
  et~al.}{2003}]{Castaneda/DiazGimenez/RiosRull:2003}
Casta{\~n}eda, A., J.~D\'{i}az-Gim\'{e}nez, and J.-V. R\'{i}os-Rull (2003,
  August).
\newblock Accounting for {U.S.} earnings and wealth inequality.
\newblock {\em Journal of Political Economy\/}~{\em 111\/}(4), 818--857.

\bibitem[\protect\citeauthoryear{Davies, Sandstr\"{o}m, Shorrocks, and
  Wolff}{Davies et~al.}{2011}]{Davies/Sandstrom/Shorrocks/Wolff:2011}
Davies, J.~B., S.~Sandstr\"{o}m, A.~Shorrocks, and E.~N. Wolff (2011, March).
\newblock The level and distribution of global household wealth.
\newblock {\em Economic Journal\/}~{\em 121\/}(551), 223--254.

\bibitem[\protect\citeauthoryear{D\'{i}az-Gim\'{e}nez, Glover, and
  R\'{i}os-Rull}{D\'{i}az-Gim\'{e}nez
  et~al.}{2011}]{DiazGimenez/Glover/RiosRull:2011}
D\'{i}az-Gim\'{e}nez, J., A.~Glover, and J.-V. R\'{i}os-Rull (2011, February).
\newblock Facts on the {U.S.} distributions of earnings, income, and wealth in
  the united states: 2007 update.
\newblock {\em Federal Reserve Bank of Minneapolis Quarterly Review\/}~{\em
  34\/}(1), 2--31.

\bibitem[\protect\citeauthoryear{Duffie}{Duffie}{2001}]{Duffie:2001}
Duffie, D. (2001).
\newblock {\em Dynamic Asset Pricing Theory}.
\newblock Princeton, NJ: Princeton University Press.

\bibitem[\protect\citeauthoryear{Fargione, Lehman, and Polasky}{Fargione
  et~al.}{2011}]{Fargione/Lehman/Polasky:2011}
Fargione, J.~E., C.~Lehman, and S.~Polasky (2011, July).
\newblock Entrepreneurs, chance, and the deterministic concentration of wealth.
\newblock {\em PloS ONE\/}~{\em 6\/}(7), e20728.

\bibitem[\protect\citeauthoryear{Fernholz}{Fernholz}{2002}]{Fernholz:2002}
Fernholz, E.~R. (2002).
\newblock {\em {Stochastic Portfolio Theory}}.
\newblock New York, NY: Springer-Verlag.

\bibitem[\protect\citeauthoryear{Fernholz}{Fernholz}{2015}]{Fernholz:2015a}
Fernholz, R.~T. (2015, July).
\newblock A model of economic mobility and the distribution of wealth.
\newblock \emph{mimeo, Claremont McKenna College}.

\bibitem[\protect\citeauthoryear{Fernholz and Fernholz}{Fernholz and
  Fernholz}{2014}]{Fernholz/Fernholz:2014}
Fernholz, R.~T. and R.~Fernholz (2014, July).
\newblock Instability and concentration in the distribution of wealth.
\newblock {\em Journal of Economic Dynamics and Control\/}~{\em 44}, 251--269.

\bibitem[\protect\citeauthoryear{Flavin and Yamashita}{Flavin and
  Yamashita}{2002}]{Flavin/Yamashita:2002}
Flavin, M. and T.~Yamashita (2002, March).
\newblock Owner-occupied housing and the composition of the household
  portfolio.
\newblock {\em American Economic Review\/}~{\em 92\/}(1), 345--362.

\bibitem[\protect\citeauthoryear{Gabaix}{Gabaix}{1999}]{Gabaix:1999}
Gabaix, X. (1999, August).
\newblock Zipf's law for cities: An explanation.
\newblock {\em Quarterly Journal of Economics\/}~{\em 114\/}(3), 739--767.

\bibitem[\protect\citeauthoryear{Gabaix}{Gabaix}{2009}]{Gabaix:2009}
Gabaix, X. (2009, 05).
\newblock Power laws in economics and finance.
\newblock {\em Annual Review of Economics\/}~{\em 1\/}(1), 255--294.

\bibitem[\protect\citeauthoryear{Guvenen}{Guvenen}{2007}]{Guvenen:2007}
Guvenen, F. (2007, June).
\newblock Learning your earning: Are labor income shocks really very
  persistent?
\newblock {\em American Economic Review\/}~{\em 97\/}(3), 687--712.

\bibitem[\protect\citeauthoryear{Guvenen}{Guvenen}{2009}]{Guvenen:2009}
Guvenen, F. (2009, January).
\newblock An empirical investigation of labor income processes.
\newblock {\em Review of Economic Dynamics\/}~{\em 12\/}(1), 58--79.

\bibitem[\protect\citeauthoryear{Guvenen, Karahan, Ozkan, and Song}{Guvenen
  et~al.}{2015}]{Guvenen/Karahan/Ozkan/Song:2015}
Guvenen, F., F.~Karahan, S.~Ozkan, and J.~Song (2015, March).
\newblock What do data on millions of {U.S.} workers reveal about life-cycle
  earnings risk?
\newblock \emph{mimeo, University of Minnesota}.

\bibitem[\protect\citeauthoryear{Jones}{Jones}{2014}]{Jones:2014}
Jones, C.~I. (2014).
\newblock Pareto and piketty: The macroeconomics of top income and wealth
  inequality.
\newblock Forthcoming in \emph{Journal of Economic Perspectives}.

\bibitem[\protect\citeauthoryear{Jones and Kim}{Jones and
  Kim}{2014}]{Jones/Kim:2014}
Jones, C.~I. and J.~Kim (2014, October).
\newblock A schumpeterian model of top income inequality.
\newblock \emph{mimeo, Stanford GSB}.

\bibitem[\protect\citeauthoryear{Karatzas and Shreve}{Karatzas and
  Shreve}{1991}]{Karatzas/Shreve:1991}
Karatzas, I. and S.~E. Shreve (1991).
\newblock {\em {Brownian Motion and Stochastic Calculus}}.
\newblock New York, NY: Springer-Verlag.

\bibitem[\protect\citeauthoryear{Karatzas and Shreve}{Karatzas and
  Shreve}{1998}]{Karatzas/Shreve:1998}
Karatzas, I. and S.~E. Shreve (1998).
\newblock {\em {Methods of Mathematical Finance}}.
\newblock New York, NY: Springer-Verlag.

\bibitem[\protect\citeauthoryear{Krussel and Smith}{Krussel and
  Smith}{1998}]{Krussel/Smith:1998}
Krussel, P. and A.~A. Smith (1998, October).
\newblock Income and wealth heterogeneity in the macroeconomy.
\newblock {\em Journal of Political Economy\/}~{\em 106\/}(5), 867--896.

\bibitem[\protect\citeauthoryear{Moskowitz and Vissing-Jorgensen}{Moskowitz and
  Vissing-Jorgensen}{2002}]{Moskowitz/Vissing-Jorgensen:2002}
Moskowitz, T.~J. and A.~Vissing-Jorgensen (2002, September).
\newblock The returns to entrepreneurial investment: A private equity premium
  puzzle?
\newblock {\em American Economic Review\/}~{\em 92\/}(4), 745--778.

\bibitem[\protect\citeauthoryear{Nielsen}{Nielsen}{1999}]{Nielsen:1999}
Nielsen, L.~T. (1999).
\newblock {\em Pricing and Hedging of Derivative Securities}.
\newblock New York, NY: Oxford University Press.

\bibitem[\protect\citeauthoryear{Nirei}{Nirei}{2009}]{Nirei:2009}
Nirei, M. (2009, July).
\newblock Pareto distributions in economic growth models.
\newblock \emph{Institute of Innovation Research Working Paper 09-05}.

\bibitem[\protect\citeauthoryear{Piketty}{Piketty}{2014}]{Piketty:2014}
Piketty, T. (2014).
\newblock {\em Capital in the Twenty-First Century}.
\newblock Cambridge, MA: Harvard University Press.

\bibitem[\protect\citeauthoryear{Saez and Zucman}{Saez and
  Zucman}{2014}]{Saez/Zucman:2014}
Saez, E. and G.~Zucman (2014, October).
\newblock Wealth inequality in the {United States} since 1913: Evidence from
  capitalized income tax data.
\newblock \emph{NBER Working Paper 20625}.

\bibitem[\protect\citeauthoryear{Wolff}{Wolff}{2010}]{Wolff:2010}
Wolff, E.~N. (2010, March).
\newblock Recent trends in household wealth in the united states: Rising debt
  and the middle-class squeeze---an update to 2007.
\newblock \emph{Levy Economics Institute of Bard College Working Paper No.
  589}.

\end{thebibliography}

\end{spacing}

\begin{table}[p]
\vspace{5pt}
\begin{center}
\setlength{\extrarowheight}{3pt}
\begin{tabular} {|c|cc|}

\hline

Household Wealth & Low Estimate & High Estimate  \\
Percent Rank  & Volatility $\s_k$ & Volatility $\s_k$ \\

\hline

0-10     & 0.283  & 0.286  \\
10-20   & 0.283  & 0.294  \\
20-40   & 0.283  & 0.316  \\
40-60   & 0.283  & 0.392  \\
60-100 & 0.283  & 1.662  \\

\hline

\end{tabular}
\end{center}
\vspace{-7pt} \caption{Low and high estimates of the volatilities $\s_k$.} \label{sigmasTab}
\end{table}


\begin{table}[p]
\vspace{5pt}
\begin{center}
\setlength{\extrarowheight}{3pt}
\begin{tabular} {|c|cc|}

\hline

Household Wealth           & Wealth Shares with & Wealth Shares with  \\
Percent Rank  & Low Estimate Volatility $\s_k$ & High  Estimate Volatility $\s_k$ \\

\hline

0-0.01    & 11.1\%  & 11.1\%   \\
0.01-0.1 & 10.8\%  & 10.8\%  \\
0.1-0.5   & 12.4\%  & 12.4\%    \\
0.5-1      & 7.2\%    & 7.2\%     \\
1-10       & 35.7\%  & 35.7\%  \\
10-100   & 22.8\%  & 22.8\%   \\

\hline

\end{tabular}
\end{center}
\vspace{-7pt} \caption{Household wealth shares for different estimates of the volatilities $\s_k$ under Scenario 1, which assumes that the 2012 U.S. wealth distribution is stable.} \label{futureSharesTab1}
\end{table}


\begin{table}[p]
\vspace{5pt}
\begin{center}
\setlength{\extrarowheight}{3pt}
\begin{tabular} {|c|cc|}

\hline

Household Wealth           & Wealth Shares with & Wealth Shares with  \\
Percent Rank  & Low Estimate Volatility $\s_k$ & High Estimate Volatility $\s_k$  \\

\hline

0-0.01    & 36.8\%  & 35.9\%    \\
0.01-0.1 & 7.9\%   & 8.1\%       \\
0.1-0.5   & 8.2\%  & 8.5\%    \\
0.5-1      & 4.7\%    & 4.9\%     \\
1-10       & 23.4\%  & 24.2\%   \\
10-100   & 19.0\%  & 18.5\%    \\

\hline

\end{tabular}
\end{center}
\vspace{-7pt} \caption{Household wealth shares for different estimates of the volatilities $\s_k$ under Scenario 2, which assumes that the share of wealth held by the top 0.01\% of households is increasing by 1\% per year while the share of total wealth held by the bottom 90\% of households is decreasing by 0.5\% per year.} \label{futureSharesTab2}
\end{table}


\begin{table}[p]
\vspace{5pt}
\begin{center}
\setlength{\extrarowheight}{3pt}
\begin{tabular} {|c|cc|}

\hline

Household Wealth           & Wealth Shares with & Wealth Shares with  \\
Percent Rank  & Low Estimate Volatility $\s_k$ & High Estimate Volatility $\s_k$ \\

\hline

0-0.01    & 87.9\% & 85.9\%  \\
0.01-0.1 & 2.0\%   & 2.3\%    \\
0.1-0.5   & 1.5\%   & 1.8\%     \\
0.5-1      & 0.8\%   & 1.0\%     \\
1-10       & 3.9\%   & 4.8\%    \\
10-100   & 3.9\%   & 4.2\%    \\

\hline

\end{tabular}
\end{center}
\vspace{-7pt} \caption{Household wealth shares for different estimates of the volatilities $\s_k$ under Scenario 3, which assumes that the shares of wealth held by the top 0.01\% and 0.01-0.1\% of households are increasing by 1.5\% and 0.5\% per year, respectively, while the share of total wealth held by the bottom 90\% of households is decreasing by 1\% per year.} \label{futureSharesTab3}
\end{table}


\begin{table}[p]
\vspace{5pt}
\begin{center}
\setlength{\extrarowheight}{3pt}
\begin{tabular} {|c|cc|}

\hline

Household Wealth           & Wealth Shares with & Wealth Shares with \\
Percent Rank  & Low Estimate Volatility $\s_k$ & High Estimate Volatility $\s_k$  \\

\hline

0-0.01    & 100\% & 100\%   \\
0.01-0.1 & 0\%    & 0\%       \\
0.1-0.5   & 0\%    & 0\%       \\
0.5-1      & 0\%    & 0\%       \\
1-10       & 0\%   & 0\%        \\
10-100   & 0\%   & 0\%         \\

\hline

\end{tabular}
\end{center}
\vspace{-7pt} \caption{Household wealth shares for different estimates of the volatilities $\s_k$ under Scenario 4, which assumes that the shares of wealth held by the top 0.01\% and 0.01-0.1\% of households are increasing by 3\% and 1\% per year, respectively, while the share of total wealth held by the bottom 90\% of households is decreasing by 1.5\% per year.} \label{futureSharesTab4}
\end{table}


\begin{table}[p]
\vspace{5pt}
\begin{center}
\setlength{\extrarowheight}{3pt}
\begin{tabular} {|c|cc|}

\hline

Household Wealth           & Wealth Shares with & Wealth Shares with \\
Percent Rank  & Low Estimate Volatility $\s_k$ & High Estimate Volatility $\s_k$  \\

\hline

0-0.01    & 1.5\%   & 1.5\%     \\
0.01-0.1 & 4.0\%   & 4.1\%     \\
0.1-0.5   & 8.4\%   & 8.4\%      \\
0.5-1      & 6.7\%    & 6.8\%     \\
1-10       & 44.9\%  & 45.0\%  \\
10-100   & 34.6\%  & 34.2\%   \\

\hline

\end{tabular}
\end{center}
\vspace{-7pt} \caption{Household wealth shares with a 1-2\% progressive capital tax on the top 1\% of households for different estimates of the volatilities $\s_k$ under Scenario 1.} \label{taxSharesTab1}
\end{table}


\begin{table}[p]
\vspace{5pt}
\begin{center}
\setlength{\extrarowheight}{3pt}
\begin{tabular} {|c|cc|}

\hline

Household Wealth           & Wealth Shares with & Wealth Shares with  \\
Percent Rank  & Low Estimate Volatility $\s_k$ & High Estimate Volatility $\s_k$  \\

\hline

0-0.01    & 1.8\% & 1.9\%        \\
0.01-0.1 & 3.8\%   & 3.9\%      \\
0.1-0.5   & 7.7\%   & 7.9\%       \\
0.5-1      & 6.2\%   & 6.3\%       \\
1-10       & 41.0\%   & 42.0\%    \\
10-100   & 39.6\%   & 37.9\%     \\

\hline

\end{tabular}
\end{center}
\vspace{-7pt} \caption{Household wealth shares with a 1-2\% progressive capital tax on the top 1\% of households for different estimates of the volatilities $\s_k$ under Scenario 2.} \label{taxSharesTab2}
\vspace{30pt}
\end{table}


\begin{table}[p]
\vspace{5pt}
\begin{center}
\setlength{\extrarowheight}{3pt}
\begin{tabular} {|c|cc|}

\hline

Household Wealth           & Wealth Shares with & Wealth Shares with \\
Percent Rank  & Low Estimate Volatility $\s_k$ & High Estimate Volatility $\s_k$  \\

\hline

0-0.01    & 2.4\% & 2.5\%        \\
0.01-0.1 & 3.9\%   & 4.1\%      \\
0.1-0.5   & 7.2\%   & 7.6\%       \\
0.5-1      & 5.7\%   & 6.0\%       \\
1-10       & 37.6\%   & 39.2\%    \\
10-100   & 43.3\%   & 40.7\%     \\

\hline

\end{tabular}
\end{center}
\vspace{-7pt} \caption{Household wealth shares with a 1-2\% progressive capital tax on the top 1\% of households for different estimates of the volatilities $\s_k$ under Scenario 3.} \label{taxSharesTab3}
\end{table}


\begin{table}[p]
\vspace{5pt}
\begin{center}
\setlength{\extrarowheight}{3pt}
\begin{tabular} {|c|cc|}

\hline

Household  Wealth         & Wealth Shares with & Wealth Shares with  \\
Percent Rank                & Low Estimate Volatility $\s_k$ & High Estimate Volatility $\s_k$ \\

\hline

0-0.01    & 14.6\%    & 14.8\%       \\
0.01-0.1 & 3.8\%    & 4.0\%       \\
0.1-0.5   & 6.0\%    & 6.4\%        \\
0.5-1      & 4.7\%    & 4.9\%        \\
1-10       & 30.5\%   & 32.3\%    \\
10-100   & 40.5\%   & 37.5\%     \\

\hline

\end{tabular}
\end{center}
\vspace{-7pt} \caption{Household wealth shares with a 1-2\% progressive capital tax on the top 1\% of households for different estimates of the volatilities $\s_k$ under Scenario 4.} \label{taxSharesTab4}
\vspace{170pt}
\end{table}









\end{document}